\documentclass[10pt]{article}
\usepackage{makeidx}
\usepackage{psfig}
\usepackage{epsf}
\usepackage{graphicx}

\usepackage{latexsym}
\usepackage{amssymb}
\usepackage{changebar}
\usepackage{booktabs}

\makeindex

\begin{document}

%%%%%%%%%%%%%%%%%%%%%% Heikos Kommandos %%%%%%%%%%%%%%%%%%%%%%%
\setlength{\changebarsep}{-33pt}

\newcommand{\graph} {\ensuremath{G}}
\newcommand{\Aset}  {\ensuremath{\mathit{E}}}
\newcommand{\Uset} {\ensuremath{\mathit{Y}}}
\newcommand{\Vset}  {\ensuremath{\mathit{X}}}
\newcommand{\qed}{\hfill\ensuremath{\rm QED}}
\newcommand{\bipartite} {\ensuremath{B}}
\newcommand{\s}{s}
\renewcommand{\t}{t}
\renewcommand{\ss}{\sigma}

\newcommand{\be}{\begin{equation}}
\newcommand{\ee}{\end{equation}}
\newcommand{\ba}{\begin{array}}
\newcommand{\ea}{\end{array}}
\newcommand{\bc}{\begin{center}}
\newcommand{\ec}{\end{center}}

\newcommand{\spacea}{\rule{0.5cm}{0mm}}
\newcommand{\spaceb}{\rule{1.0cm}{0mm}}
\newcommand{\spacec}{\rule{1.5cm}{0mm}}
\newcommand{\spaced}{\rule{2.0cm}{0mm}}
\newcommand{\spacee}{\rule{2.5cm}{0mm}}

\newcommand{\matching} {\ensuremath{\mathit{M}}}
\newcommand{\aug} {\ensuremath{\mathit{A}_p}}
\newcommand{\blossom} {\ensuremath{\mathit{B}}}
\newcommand{\weight}{\ensuremath{w}}
\newcommand{\reals}{\ensuremath{\Re}}
\newcommand{\order}[1] {\ensuremath{\mathcal{O}(#1)}}
%%%%%%%%%%%%%%%%%%%%%%%%%%%%% ENDE Heiko Kommandos %%%%%%%%%%%%%

\newcommand{\myvec}[1]{\underline{#1}}
\newcommand{\mymatrix}[1]{\underline{\underline{#1}}}
\newcommand{\narrowcaption}[1]{
\begin{minipage}{0.85\textwidth}
\caption{#1}
\end{minipage}}

\newlength{\boxlength}
\setlength{\boxlength}{\textwidth}
\addtolength{\boxlength}{-2\parindent}

\newenvironment{example}[1]
{\rule{\textwidth}{0.5mm}\vspace{-0.1cm}\begin{quote}
\underline{Example}: #1\\[0.1cm]}
{\hfill$\Box$\\
\end{quote}\vspace*{-0.7cm}\rule{\textwidth}{0.5mm}\\[0.0cm]}

\newenvironment{exampleNOL}[1]
{\rule{\textwidth}{0.5mm}\vspace{-0.1cm}\begin{quote}
\underline{Example}: #1\\[0.1cm]}
{\hfill$\Box$\\
\end{quote}}

\newenvironment{algorithm}[1]
{
\begin{tabbing} xx \= xx \= xx \= xx \= xx \= xx \= xx \= xxxxx \kill
 {\bf algorithm} #1\\
 {\bf begin}\\
}
{
 {\bf end}\\
 \end{tabbing}
}

\newenvironment{procedure}[1]
{
\begin{tabbing} xx \= xx \= xx \= xx \= xx \= xx \= xx \= xxxxxxxxxxxxxxxxx \= xxx \kill
 {\bf procedure} #1\\
 {\bf begin}\\
}
{
 {\bf end}\\
 \end{tabbing}
}

\newenvironment{definition}[1]
{
{\bf Definition: #1}
}{\hfill}

\newenvironment{proof}
{ {\bf Proof:}}
{\hfill QED}

% for primary index
\newcommand{\ii}[1]{{\it #1}}

\title{A practical guide to computer simulations \footnote{Taken from
the book: A.K. Hartmann and H. Rieger, {\em Optimization Algorithms in
Physics}, (Wiley-VCH, Berlin, Weinheim 2001), ISBN  3-527-40307-8,
with permission of Wiley-VCH, see http://www.wiley.com. 
This document may be distributed freely
in electronic and non-electronic form, provided that no changes are
performed to it.}}
\author{Alexander K. Hartmann\\
{\small University of G\"ottingen, Germany}\\
{\small hartmann@theorie.physik.uni-goettingen.de} \\[5mm]
Heiko Rieger\\
{\small University of Saarbr\"ucken, Germany} \\
{\small rieger@lusi.uni-sb.de}}
\maketitle
\begin{abstract}
Here practical aspects of conducting research via computer simulations are
discussed. The following issues are addressed: software engineering,
object-oriented software development, programming style, macros, {\em
make files}, scripts, libraries, random numbers, testing, debugging,
data plotting, curve fitting, finite-size scaling, information
retrieval, and preparing presentations.

Because of the limited space, usually only short introductions to the
specific areas are given and references to more extensive literature
are cited. All examples of code are in C/C++.
\end{abstract}

\tableofcontents
\label{chap-practical}

Here practical aspects of conducting research via computer simulations are
discussed. It is assumed that you are familiar with an operating
system such as UNIX (e.g.\ Linux), \index{UNIX} \index{Linux} a high-level
programming language such as C, \index{C programming language} 
Fortran \index{Fortran}
or Pascal \index{Pascal} and have some experience
with at least small software projects. 

Because of the limited space, usually only short introductions to the
specific areas are given and references to more extensive literature
are cited. All examples of code are in C/C++.

First, a short introduction to software engineering is
given and several hints allowing the construction of efficient and reliable
code are stated. In the second section a short introduction to
object-oriented software development is presented. In particular, it is shown
that this kind of programming style can be achieved with standard
procedural languages such as C as well. Next, practical hints concerning
the actual process of writing the code are given. In the fourth
section macros are introduced. Then it is shown how the development
of larger pieces of code can be organized with the help of so called 
{\em make files}\/. In the
subsequent section the benefit of using libraries like {\em Numerical
Recipes}\/ or {\em LEDA}\/ are explained and
it is shown how you can build your own libraries. In the sixth
section the generation of random numbers is covered while in the eighth
section three very useful debugging tools are presented. Afterwards,
programs to perform data analysis, curve fitting and finite-size
scaling are explained. In the last section an introduction to
information retrieval and literature search in the Internet and to the
preparation of presentations and publications is given.

\section{Software Engineering}
\label{sec-engineering}
\index{software!engineering|(}

When you are creating a program, you should never just
start writing the code. In this way only tiny software projects
such as scripts \index{script}
can be completed successfully. Otherwise your code will
probably be very inflexible and contain several hidden errors which are
very hard to find. If several people are involved in a project, it is
obvious that a considerable amount of planning is necessary.

But even when you are programming alone, which is not unusual in physics, 
the first step you should undertake is just to sit down and
think for a while. This will save you a lot of time and effort later on. To
emphasize the  need for structuring in the software development
process,  the art
of writing good programs is usually called {\em software
  engineering}\/. There are many specialized books in this fields, see e.g.
Refs. \cite{PRA-sommerville1989,PRA-ghezzi1991}. 
Here just the steps that should be undertaken to create a
 sophisticated software
development \index{software!development}
process are stated. The following descriptions refer to the
usual situation you find in physics: one or a few people are involved
in the project. How to manage the development of big programs
involving many developers is explained in literature.

\begin{itemize}
\item {\bf Definition of the problem and solution strategies\\} 
You should write down which
problem you would like to solve. 
Drawing diagrams is always helpful!
Discuss your problem with others and tell them how you would like to
solve it. In this context many questions may
appear, here some examples are given:
\begin{itemize}
\item What is the input\index{input}
 you have to supply? In  case you have only a few
  parameters, \index{parameters}
they can be passed to the program via options. In other
  cases, especially when chemical systems are to be simulated, many
  parameters have to be controlled and it may be advisable to use extra
  parameter files.
\item Which results do you want to obtain and which quantities do you
  have to analyze? Very often it is useful to write the raw results
  of your simulations, e.g.\ the positions of all atoms or the
  orientations of all spins of your system, to a configuration
  file. \index{configuration file} The physical results can be obtained by
  post-processing. Then, in case new questions arise, it is very easy
  to analyze the data again. When using configuration files, you
  should estimate the amount of data you generate. Is there enough
  space on your disk? It may be helpful, to include the compression of
  the data files directly in your programs\footnote{In C this 
can be achieved by
    calling {\tt system("gzip -f }{\it $<$filename$>$}"); after the file has
      been written and closed.}.
\item Can you identify ``objects'' \index{object}
in your problem? Objects may be
  physical entities like atoms or molecules, but also internal
  structures like nodes in a tree or elements of tables. Seeing the
  system and the program as a hierarchical collection of objects usually makes
  the problem easier to understand.
More on object-oriented development can be found in Sec.
  \ref{sec-oo}. 
\item Is the program to be extended later on? Usually a code is ``never''
  finished. You should foresee later extensions of the
  program and set up everything in a way it can be reused easily.
\item Do you have existing programs available which can be included into the
  software project? If you have implemented your previous projects in the
  above mentioned fashion, it is very likely that you can recycle some
  code. But this requires experience and is not very easy to achieve at
  the beginning. But over the years you will have a growing library
  of programs which enables you to finish future software projects much
  quicker.

Has somebody else  created a program which you can
  reuse? \index{software!reuse}
  Sometimes you can rely on external code like libraries. Examples are
  the {\em Numerical Recipes}\/ \index{Numerical Recipes}
\cite{PRA-numrec1995} and the {\em LEDA}\/ library
  \index{library}
\index{LEDA library} \cite{PRA-leda1999} which are covered in Sec.\ \ref{sec-lib}.
\item Which algorithms are known? Are you sure that you can solve the
  problem at all? 
%By reading this book you should have
% a good knowledge of the basic methods which are available. 
%  Naturally, here only a small subset can be presented. 
Many other techniques have been invented already.
You should always search the literature for
  solutions which already exist. 
How searches can be simplified by using
  electronic data bases is covered more deeply in
  Sec.\ \ref{sec-literature}.

  Sometimes it is necessary to invent new
  methods. This part of a project may be the most time
  consuming. 
\end{itemize}
\item {\bf Designing data structures\\}\index{data!structures}
Once you have identified the basic objects in your systems, you have
to think about how to represent them in the code. Sometimes it is
sufficient to define some {\em struct}\/ \index{struct}
types in C (or simple {\em classes}\/ \index{class}
in C++). But usually you
will need to design a large set of data structures, referencing each
other in a complicated way. 

A sophisticated design of the data structures will lead to 
a better organized program, usually it will even run
faster. For example, consider a set of vertices of a
graph. \index{graph}
Then assume that you have several lists \index{list}
$L_i$ each containing elements
referencing the vertices
of degree $i$. When the graph is altered in your program and thus the
degrees \index{vertex!degree} 
of the vertices \index{vertex} change, it is sometimes necessary to remove 
a vertex from one
list and insert it into another. In this case you will gain speed,
when your vertices data structures 
also contain pointers to the positions where they
are stored in the lists. Hence, removing and inserting vertices in the
lists will take only a constant amount of time. Without these additional
pointers, the insert and delete operations have to scan partially through the
lists to locate the elements, leading to a linear time complexity of
these operations.

Again, you should perform the design of the data structures in a way,
that later extensions are facilitated. For example when treating lattices
of Ising spins, you should use data structures which are independent of the
dimension or even of the structure of the lattice, an example is given 
in Sec.\ \ref{sec-macros}.

When you are using external libraries, \index{library}
usually they have some data
types included. The above mentioned LEDA \index{LEDA library}
library  has many predefined
data types like arrays, \index{array} stacks, \index{stack} lists
\index{list} or graphs. \index{graph} You can have
e.g.\ arrays of arbitrary objects, for example arrays of strings. Furthermore,
it is possible to combine the data types
in complicated ways, e.g.\ you can define a stack of graphs having
strings attached to the vertices.

\item {\bf Defining small tasks\\} \index{small tasks} \index{tasks}
After setting up the basic data types, you should think about which
basic and complex operations, i.e.\ which subroutines,
 you need to manipulate the objects of your
simulation. Since you have already thought a lot about your problem,
you have a good overview, which operations \index{operation} may occur. 
You should break down the final task ``perform simulation'' 
into small subtasks, this means you use a {\em top down}\/ approach
\index{top down approach} in the design process. 
It is not possible to write a program
in a sequential way as one code. For the actual implementation, 
a {\em bottom up}\/ \index{bottom up approach} approach 
is recommended. This means you should start with the most basic
operations. Later on you can use them to create more complicated
operations. 
 As always, you should define the subroutines in a way that
they can be applied in a flexible way and extensions are easy to
perform.

But it is not necessary that you must identify all basic operations at the
beginning. During the development of the code, new applications may
arise, which lead to the need for further operations. Also it may be
required to change or extend the data structures defined before. However,
the more you think in advance, the less you need to change the program
later on.

As an example, the problem of finding ground states in Ising spin
glasses \index{spin glass}
via simulated annealing \index{simulated annealing}
is considered. Some of basic operations \index{basic operations}
\index{operation!basic} are: 
\begin{itemize}
\item Set up the data structures for storing the realizations of
  the interactions and for storing the spin glass configurations.
\item Create a random realization of the interactions.
\item Initialize a random spin configuration.
\item Calculate the energy of a spin in the local field of its neighbors.
\item Calculate the total energy of a system.
\item Calculate the energy changes associated with a spin flip.
\item Execute  a Monte Carlo step.
\item Execute  a whole annealing run.
\item Calculate the magnetization.
\item Save a realization and corresponding spin configurations in a file.
\end{itemize}
  
It is not necessary to define a corresponding
subroutine for all operations.
 Sometimes they require only a few numbers of lines in the
code, like the calculation of the energy of one spin in the
example above.
 In this case, such operations can be written directly in the code,
 or a macro \index{macro} (see Sec.\ \ref{sec-macros}) can be used.

\item {\bf Distributing work\\} \index{distributing work}
In case several people are involved in a project, the next step is to
split up the work between the coworkers. If several types of objects
appear in the program design, 
a natural approach is to make everyone responsible for one or
several types of objects and the related operations. The code should
be broken up into several modules (i.e.\ source files), such that every
module is written by only one person. This makes the
implementation easer and also helps testing the code (see below). 
Nevertheless, the partitioning of the work
requires much care, since quite often some modules or data types
depend on others. For this reason, the actual implementation of a data
type should be hidden. This means that all interactions should be
performed through
exactly defined interfaces which do not depend on the internal
representation, see also Sec.\ \ref{sec-oo} on object-oriented programming.

When several people are editing the same files, which is usually
necessary later on, even when initially each file was created by only
one person, then you should use a {\em source-code management system}\/.
 \index{source-code management system} It prevents
 several people from performing changes on the same file in
parallel, which would cause a lot of trouble. 
Additionally, a source-code management system 
enables you to keep track of all changes made.
An example of such a system is the {\em Revision Control System}\/
\index{Revision Control System}(RCS), \index{RCS}
which is freely available through the {\em GNU project}\/ \cite{PRA-gnu}
\index{GNU project} and part of the free operating system {\em Linux}\/.
\index{Linux}

\item {\bf Implementing the code\\} \index{implementation}
With  good preparation, the actual implementation becomes only a
small part of the software development process. General style
rules, guaranteeing clear structured code, which can even be understood
several months later, are explained in Sec.\ \ref{sec-style}.
You should use  a different file, i.e.\ a different module, 
for each coherent unit of data structures and subroutines;
when using an object oriented language you should define different
classes (see Sec.\ \ref{sec-oo}). \index{class} This rule should be obeyed
 for the case of a  one-person project as well. Large software projects
containing many modules are
easily maintained via {\em makefiles}\/ (see Sec.\ \ref{sec-make}).
\index{makefile@{\em{}makefile}}

Each subroutine and each module \index{module}
should be tested \index{testing}
separately, before
integrating many modules into one program. In the following some
general hints concerning  testing are presented.
\item {\bf Testing\\}
When performing tests on single subroutines, standard cases
 usually are used. This is the reason why many errors become apparent much
later. Then, because the modules have already been integrated into one single
program, errors are much harder to localize. For this reason, you should
always try to find special and rare cases as well when testing a
subroutine. Consider for example a procedure which inserts an element
into a list. Then not only inserting in the middle of the list, but
also at the beginning, at the end and into an empty list must be
tested. Also, it is strongly recommended to read 
your code carefully once again before considering it finished. 
In this way many bugs can be found easily
which otherwise must be tracked down by intensive debugging.

The actual debugging \index{debugging}
of the code can be performed by placing print
instructions at selected positions in the code. But this approach is
quite time consuming, because you have to modify and recompile your
program several times. Therefore, it is advisable to use debugging tools
like a {\em source-code debugger}\/ \index{source-code debugger}
 and a program for checking the memory
management. More about these tools can be found in
Sec.\ \ref{sec-testing}. But usually you also need  special operations
 which are not covered by an available tool.
You should always write a procedure which prints out the
current instance of the system that is simulated, e.g.\ the nodes and
edges of a graph or the interaction constants of an Ising system. This
facilitates the types of tests, which are described in the following.

After the raw operation of the subroutines has been verified, more
complex tests can be performed. When e.g.\ testing an optimization
routine, you should compare the outcome of the calculation for a
 small system with the result which can be obtained by
 hand. If the outcome is different from the expected result, the small
 size of the test system allows you to follow the execution of the program
 step by step. For each operation you should think about the expected
 outcome and compare it with the result originating from the running program.

Furthermore, it is very useful to compare the outcome of  different methods
 applied to the same problem.  For example, you know
 that there must be something wrong, in  case an approximation method
 finds a better value than your ``exact'' algorithm. Sometimes 
 analytical solutions are  available, at least for special
 cases. Another approach is to use invariants. For example, when
 performing a Molecular Dynamics simulation 
\index{Molecular Dynamics simulation}\index{simulation!Molecular Dynamics} 
of an atomic/molecular
 system (or a galaxy), 
 energy and momentum must be conserved; \index{energy!conservation}
\index{conservation!of energy} \index{momentum conservation}
\index{conservation!of momentum} only
 numerical rounding errors should appear. These quantities can be
 recorded very easily. If they change in time there must be a
 bug in your code. In this case, usually the formulas for the energy and the
 force are not compatible or the integration subroutine has a bug.

You should test each procedure, directly after writing it. Many developers
have experienced that the larger the interval between
implementation and tests is, the lower the motivation  becomes
for performing 
tests, resulting in more undetected bugs.

The final stage of the testing process occurs when several modules
are integrated into one large running program. In the case where you are
writing the code alone, not many surprises should appear, if you have
performed many tests on the single modules. If several
people are involved in the project, at this stage many errors
occur. But in any case, you should always remember: there is probably
no program, unless very small, which is bug free. You should know the
following important result from theoretical computer science 
\cite{PRA-lewis1981}: 
it is impossible
to invent a general method, which can prove automatically that a given
program obeys a given specification. Thus, 
all tests must be designed to match the current code.

In case a program is changed or extended several times, you should
always keep the old versions, because it is quite common that by
editing new bugs are introduced. In that case, you can compare your
new code with the older version. Please note that editors like emacs
only keep the second latest version as backup, so you have to take care
of this problem yourself unless you use a source-code management
system, where  you are lucky, because it keeps all older version automatically.

For C programmers, it is always advisable to apply the {\tt -Wall}
\index{Wall option@{\tt{}-Wall} option}
(warning level: all) option. Then several bugs already show up during
the compiling process, for example the common mistake to use '=' in
comparisons instead of '==', or the access to uninitialized
variables\footnote{But this is not true for some C++ compilers when
  combining with option {\tt -g}.}.

In C++, some bugs can be detected by defining variables or
parameter as {\tt const}, \index{const@{\tt{}const}}
when they are considered to stay unchanged
in a block of code or subroutine. Here again, already the compiler
will complain, if attempts to alter the value of such a variable are tried.

This part finishes with a  warning: never try to save time when
performing tests. Bugs which appear later on are much much harder to
find and you will have to spend much more time than you have ``saved'' before.

\item {\bf Writing documentation\\}\index{documentation}
This part of the software development process is very often disregarded,
especially in the context of scientific research, 
where no direct customers exist. But even if you are
using your own code, you should write  good documentation. It should
consist of at least three parts:

\begin{itemize}
\item {\em Comments in the source code}\/: \index{comment}
You should place comments at the
  beginning of each module, in front of each subroutine or each
  self-defined data structure, for blocks of the code and for selected
  lines. Additionally, meaningful names for the variables are
  crucial. Following these rules makes later changes and extension of
  the program much more straightforward. 
You will find in more hints
 on how a good programming style can be achieved Sec.\ \ref{sec-style}.
\item {\em On-line help}\/: You should include a short description of the
  program, its parameters and its options in the main program. It
  should be printed, when the program is called with the wrong
  number/form of the parameters, or when the option {\tt -help}
  \index{help option@{\tt{}-help} option} is
  passed. Even when you are the author of the program, after it has grown
  larger it is quite hard to remember all options and usages.
\item {\em External documentation}\/: This part of the documentation
  process is important, when you would like to make the program
  available to
  other users or when it grows really complex. Writing  good
  instructions is really a hard job. When you remember how often you
  have complained about the instructions for a video recorder or a word
  processor, you will understand why there is a high demand for good
  authors of documentation in industry.
\end{itemize}
\item {\bf Using the code\\}
Also the actual performance of the simulation usually requires careful
preparation. Several question have to be considered, for example:
\begin{itemize}
\item How long will the different runs take? You should  perform simulations of
  small systems and extrapolate to large system sizes.
\item Usually you have to average over different runs or over several
  realizations of the disorder. The system sizes should also be chosen
  in a way that the number of samples is large enough to reduce the
  statistical fluctuations. It is better to have a reliable result for
  a small system than to treat only a few instances of a large system. 
  If your model exhibits self averaging, the larger the sample, the
  less the number of samples can be. But, unfortunately, usually the
  numerical effort grows stronger than the system size, so there will
  be a maximum system size which can be treated with satisfying
  accuracy. To estimate the accuracy, 
you should always calculate the statistical error bar \index{error
  bar}
  $\sigma(A)$ for
  each quantity $A$\footnote{The error bar
 is $\sigma(A)=\sqrt{\mbox{Var}(A)/(N-1)}$, where
  $\mbox{Var}(A)=\frac{1}{N}\sum_{i=1}^N
  a_i^2-(\frac{1}{N}\sum_{i=1}^N a_i)^2 $ is the variance of the $N$
  values $a_1,\ldots,a_N$.}.

A good rule of a thumb is that each sample
  should take not more than 10 minutes. When you have many
  computers and much time available, you can attack larger problems as
  well. 
\item Where to put the results? In many cases you have to investigate
  your model for different parameters. You should organize the
  directories where you put the data and the names of the files
in such a way that even years later
  the former results can be found quickly. You should put a
  README \index{README} file in each directory, explaining what it contains.

If you want to start a sequence of several simulations, you can write
a short script, which calls your program with different parameters
within a loop.
\item Logfiles \index{logfile}
are very helpful, where during each simulation some
  information about the ongoing processes are written automatically. 
Your program should put
  its version number and the parameters which
  have been used to start the simulation in the first line of each
  logfile. This allows a reconstruction of how the results have
  been obtained.
\end{itemize}

\end{itemize}

The steps given do not  usually occur in linear order. It is quite common
that after you have written a program and performed some simulations,
you are not satisfied with the performance or new questions
arise. 
Then you start to define new problems and the program will be
extended. It may also be necessary to extend the data structures, when
e.g.\ new attributes of the simulated models have to be included.
It is also possible that a nasty bug is still hidden in the program,
which is found later on during the actual simulations and becomes obvious by
results which cannot be explained. In this case changes cannot be
circumvented either.

In other words, the software
development process is a {\em cycle}\/ \index{software!cycle}
which is traversed several times.
As a consequence, when planning your code, you should always keep this in
mind and set up everything in a flexible way, so that extensions and code
recycling can be performed easily.
\index{software!engineering|)}

\section{Object-oriented Software Development}
\label{sec-oo}
\index{object-oriented programming|(}
\index{object|(ii}

In recent years {\em object-oriented}\/ programming 
languages like C++, Smalltalk
or Eiffel became very popular. 
But, using an object-oriented language and developing the program in
an object-oriented style are not necessarily the same, although they are
compatible. For example, you can  set up your whole
project by applying object-oriented methods even when using a
traditional {\em procedural}\/ \index{procedural programming}
programming language like C, \index{C programming language} 
Pascal \index{Pascal}
or Fortran.\index{Fortran} On the other hand,
it is possible to write very traditional programs with modern
object-oriented languages. They help to organize your programs in
terms of objects, but you have the flexibility to do it in another way
as well. In general, taking an object-oriented viewpoint facilitates
the analysis of problems and the development of programs for solving the
problems. 
Introductions to object-oriented software development can be found
e.g.\ in Refs. \cite{PRA-rumbaugh1991,PRA-johnsonbaugh1994,PRA-skansjolm1997}.
Here just the main principles are explained:

\begin{itemize}
\item {\bf Objects and methods\\} \index{methods for objects}
The real world is made of {\em objects}\/ such as traffic-lights, books or
computers. You can classify different objects according to some criteria
into {\em classes}\/. \index{class|ii}
This means different chairs belong to the
class ``chairs''. The objects of many classes can have internal  {\em
  states}\/, e.g.\ a traffic-light can be red, yellow or green. The state
of a computer is much more difficult to describe. 
Furthermore, objects are useful for the environment, because other
objects interact via {\em operations}\/ with the object. You (belonging
to the class ``human'') can read the state
of a traffic light, some central computer may set the state or even
switch the traffic light off.

Similar to the real world, you can have objects in programs as
well. The internal state of an object is given by the values of the
variables describing the object. 
Also it is possible to interact with the objects by calling subroutines
(called {\em methods}\/ in this context) associated with the objects.

Objects and the related methods are seen as coherent
units. This means you define within one {\em
  class definition}\/ the way the objects look, i.e.\ the data structures,
together with the methods which access/alter the content of the objects. 
The syntax of the class definition depends on the
programming language you use. Since implementational details are not
relevant here,
the reader is referred to the literature.

When you take the viewpoint of a pure object-oriented
programmer, then all programs can be organized as 
collections of objects calling methods of each other.
This is derived from the structure the real world
has: it is a large set of interacting objects. But for writing good
programs it is as in real life, taking an orthodox position imposes
too many 
restrictions. You should take the best of both worlds, the
object-oriented and the procedural world, depending on the actual problem.
\item {\bf Data capsuling\\} \index{data!capsuling} \index{capsuling}
When using a computer, you do not care about the implementation. When
you press a key on the keyboard, you would like to see the result
on the screen. You are not interested in how the key converts your
pressing into an electrical signal, how this signal is sent to the
input ports of the chips, how the algorithm treats the signal and so on.

Similarly, a main principle of object-oriented programming is to hide the
actual implementation of the objects. Access to them  is only allowed
via  given interfaces, \index{interface} i.e.\ via methods. 
The internal data structures are hidden, this is called {\tt private} 
\index{private@{\tt private}} in C++.
The data capsuling has several advantages: 
\begin{itemize}
\item You do not have to remember the implementation of your
  objects. When using them later on, they just appear as a black box
  fulfilling some duties.
\item You can change the implementation later on without the need to
  change the rest of the program. Changes of the implementation may be useful
  e.g.\ when you want to increase the performance of the code or
  to include new features. 
\item
  Furthermore, you can have {\em flexible data
    structures}\/: \index{data!structures}
several different types of implementations may
  coexist. Which one is chosen depends on the requirements. An example
  are graphs which can be implemented via arrays, lists, hash tables or
  in other ways. In the case of sparse graphs, \index{sparse graph}
\index{graph!sparse} the list implementation
   has a better
  performance. When the graph is almost complete, the array
  representation is favorable. Then you only have to provide the basic
  access methods, such as inserting/removing/testing vertices/edges
  and iterating over them, for the different internal
  representations. Therefore, higher-level algorithms like computing a
  spanning tree can be written in a simple way to work with all
  internal implementations. When using such a class, the user just
  has to specify the representation he wants, the rest of the
  program is independent of this choice.  
\item Last but not least, 
software debugging is made easier. Since you have only defined
  ways the data can be changed, undesired side-effects become less
  common. Also the memory management can be controlled easier.
\end{itemize}
For the sake of flexibility, convenience or speed it is possible to declare
internal variables as {\tt public}. \index{public@{\tt{}public}} 
In this case they can be accessed directly from outside.

\item {\bf Inheritance\\}\item{inheritance}
This means lower level objects can be specializations of
higher level objects. For example the class of (German) ``ICE trains''
is a subclass of ``trains'' which itself is a subclass of ``vehicles''.

In computational physics, you may have a basic class of ``atoms''
containing mass, position and velocity, and built upon this 
a class of ``charged atoms'' by including the value of the charge. Then
you can use the subroutines you have written for the uncharged atoms,
like moving the particles or calculating correlation functions, 
for the charged atoms as well.

A similar form of hierarchical organization of objects works 
the other way round: higher level objects can be defined in terms of lower
level objects. For example a book is composed of many objects
belonging to the class ``page''. Each page can be regarded as
 a collection of many ``letter'' objects.

For the physical example above, when modeling chemical systems, you can
have ``atoms'' as basic objects and use them to define ``molecules''. Another
level up would be the ``system'' object, which is a collection of
molecules. 

\item {\bf Function/operator overloading\\} \index{operator overloading}
This inheritance of methods to lower level classes is an example of
{\em operator overloading}\/. It just means that you can have methods
for different classes having the same name, sometimes the same code
applies to several classes. This applies also to classes, which are
not connected by inheritance.
For example you can
define how to add integers, real numbers, complex numbers or larger
objects like lists, graphs or documents. In language like C or Pascal you
can define subroutines to add numbers and subroutines to add graphs as
well, but they must have different names. In C++ you can define the
operator ``+'' for all
different classes. Hence, the operator-overloading
mechanisms of object-oriented languages is just a tool to make the
code more readable and clearer structured. 
\item {\bf Software reuse\\}\index{software!reuse}
Once you have an idea of how to build a chair, you can do it several
times. Because you have a blueprint, the tools and the experience,
building another chair is an easy task.

This is true for building programs as well: 
both data capsuling and inheritance facilitate the reuse of
software. Once you have written your class for e.g.\ treating lists, 
you can include them in other programs as well. This is easy,
because later on you do not have to care about the implementation. With a class
designed in a flexible way, much time can be saved when realizing
new software projects.
\end{itemize}

As mentioned before, for  object-oriented programming you do not
necessarily have to use an object-oriented language. It is true that they
are helpful for the implementation and the resulting programs will look
slightly more elegant and clear, but you can program
 everything with a language like C as well.
In C an object-oriented style can be achieved very easily. As an
example a class {\tt histo} \index{class!histo@{\tt{}histo}}
 implementing histograms \index{histogram} is outlined, which
are needed for almost all types of computer simulations as evaluation
and analysis tools.

First you have to think about the data you would like to store. That
is the histogram itself, i.e.\ an array {\tt table} of bins. Each bin
just counts the number of events which fall into a small interval.
To achieve a high
degree of flexibility, the range and the number of bins must be
variable. From this, the width {\tt delta} of each bin can be
calculated. For convenience {\tt delta}  is stored as well. To count the number
of events which are outside the range of the table, 
the entries {\tt low} and {\tt high} are introduced. Furthermore,
statistical quantities like mean and variance should be available
quickly and with high accuracy. Thus, several summarized 
moments {\tt sum} of the
distribution are stored separately as well. Here the
number of moments {\tt \_HISTO\_NOM\_} is defined 
as a macro, converting this macro to
variable is straightforward.  All together, this leads to the
following C data structure: 

\begin{verbatim}

#define _HISTO_NOM_      9         /* No. of (statistical) moments */

/* holds statistical informations for a set of numbers:   */
/* histogram, # of Numbers, sum of numbers, squares, ...  */
typedef struct
{
  double           from, to;   /* range of histogram               */
  double              delta;   /* width of bins                    */  
  int                n_bask;   /* number of bins                   */
  double             *table;   /* bins                             */
  int             low, high;   /* No. of data out of range         */
  double   sum[_HISTO_NOM_];   /* sum of 1s, numbers, numbers^2 ...*/
} histo_t;
\end{verbatim}

Here, the postfix {\tt \_t} is used to stress the fact that the name 
{\tt histo\_t} denotes a type. The bins are {\tt double}
variables, which allows for more general applications.
Please note that it is still possible to access the internal
structures from outside, but it is not necessary and not
recommended. In C++, you could prevent this by declaring the
internal variables as {\tt private}. Nevertheless,
everything can be done via special subroutines. First of all
one must be able to create and delete histograms, please
note that some simple error-checking is included in the program:

\begin{verbatim}
/** creates a histo-element, where the empirical histogram **/
/** table covers the range ['from', 'to'] and is divided   **/
/** into 'n_bask' bins.                                    **/
/** RETURNS: pointer to his-Element, exit if no memory.    **/
histo_t *histo_new(double from, double to, int n_bask)
{
  histo_t *his;
  int t;

  his = (histo_t *) malloc(sizeof(histo_t));
  if(his == NULL)
  {
    fprintf(stderr, "out of memory in histo_new");
    exit(1)
  }
  if(to < from) 
  {
    double tmp;
    tmp = to; to = from; from = tmp;
    fprintf(stderr, "WARNING: exchanging from, to in histo_new\n");
  }
  his->from = from;
  his->to = to;
  if( n_bask <= 0) 
  {
    n_bask = 10;
    fprintf(stderr, "WARNING: setting n_bask=10 in histo_new()\n");
  }
  his->delta = (to-from)/(double) n_bask;
  his->n_bask = n_bask;
  his->low = 0;
  his->high = 0;
  for(t=0; t< _HISTO_NOM_ ; t++) /* initialize summarized moments */
    his->sum[t] = 0.0;
  his->table = (double *) malloc(n_bask*sizeof(double));
  if(his->table == NULL)
  {
    fprintf(stderr, "out of memory in histo_new");
    exit(1);
  }
  else
    for(t=0; t<n_bask; t++)
      his->table[t] = 0;
  }
  return(his);
}
\end{verbatim}

\begin{verbatim}
/** Deletes a histogram 'his'     **/
void histo_delete(histo_t *his)
{
  free(his->table);
  free(his);
}
\end{verbatim}

All histogram objects are created dynamically by calling
{\tt histo\_new()}, this corresponds to a call of the {\em
  constructor}\/ \index{constructor} or {\tt new} in C++. 
The objects are addressed via pointers. Whenever a method,
i.e.\ a procedure in C,
of the {\tt histo} class is called, the first argument will always be a
pointer to the corresponding histogram. This looks slightly less
elegant than writing {\tt histo.method()} in C++, but it is really the
same. When avoiding direct access, the realization using C is
perfectly equivalent to C++ or other object-oriented
languages. Inheritance can be implemented, by including pointers to
{\tt histo\_t} objects in other type definitions. When these higher
level objects are created, a call to {\tt histo\_new()} must be
included, while a call to {\tt histo\_delete()}, corresponding to the
{\em destructor}\/ \index{destructor} in C++, is necessary, to
implement a correct deletion of the more complex objects.

As a final example, the procedures for inserting an element
into the table and calculating the mean are presented. It is easy
to figure out how other subroutines for e.g.\ calculating the 
variance/higher moments or printing a histogram can be realized. The complete
library can be obtained for free \cite{PRA-histo}.

\begin{verbatim}
/** inserts a 'number' into a histogram 'his'. **/
void histo_insert(histo_t *his, double number)
{
  int t;
  double value;
  value = 1.0;
  for(t=0; t< _HISTO_NOM_; t++)
  {
    his->sum[t]+= value;;                     /* raw statistics */ 
    value *= number;
  }
  if(number < his->from)               /* insert into histogram */
    his->low++;
  else if(number > his->to)
    his->high++;
  else if(number == his->to)
    his->table[his->n_bask-1]++;
  else
    his->table[(int) floor( (number - his->from) / his->delta)]++;
}
\end{verbatim}

\newpage
\begin{verbatim}
/** RETURNS: Mean of Elements in 'his' (0.0 if his=empty) **/
double histo_mean(histo_t *his)
{
  if(his->sum[0] == 0)
    return(0.0);
  else
    return(his->sum[1] / his->sum[0]);
}
\end{verbatim}
\index{object-oriented programming|)}
\index{object|)}

\section{Programming Style}
\label{sec-style} 
\index{programming!style|(} \index{style|(}

The code should be written in a style that enables the author, and
other people as well, to understand and modify the program even years
later. Here briefly some principles  you should follow are stated. Just
a general style of description is given. Everybody is free
to choose his/her own style, as long as it is precise and consistent.

\begin{itemize}
\item Split your code into several modules. \index{module} This has several
  advantages: 
\begin{itemize}
\item When you perform changes, you have to recompile only the
  modules which have been edited. 
  Otherwise, if everything is contained in a long file, the whole
  program has to be recompiled each time again.
\item Subroutines which are related to each other can be collected in
  single modules. It is much easier to navigate in several short files
  than in one large program. 
\item After one module has been finished and tested it can be used for
  other projects. Thus, software reuse is facilitated.
\item Distributing the work among several people is impossible if
  everything is written into one file. Furthermore, you should use a
  source-code management system (see Sec.\ \ref{sec-engineering}) in
  case several people are involved in avoiding uncontrolled editing.
\end{itemize}
\item To keep your program logically structured, 
you should always put data structures and 
  implementations of  the
  operations in separate files. In C/C++ this means you have to write the data
  structures in a header ({\tt{}.h}) file \index{header file}
and the code into a source
  code ({\tt{}.c}/ {\tt{}.cpp}) file.
\item Try to find meaningful names for your variables
  \index{variable} and
  subroutines. Therefore, during the programming process it is much
  easier to remember their meanings, which helps a lot in avoiding
  bugs. Additionally, it is not necessary to look up the meaning
  frequently. For local variables like loop counters, it is sufficient
and more convenient to have short (e.g.\ one letter) names.

\begin{sloppypar}
In the beginning this might seem to take additional time (writing
e.g.\ '{\tt{}kinetic\_energy}' for a variable instead of '{\tt{}x10}').
 But several months after you have written the program, you will
 appreciate your effort, when you read the line 
\end{sloppypar}
\begin{verbatim}
kinetic_energy += 0.5*atom[i].mass*atom[i].veloc*atom[i].veloc;
\end{verbatim}
instead of
\begin{verbatim}
x10 += 0.5*x34[i].a*x34[i].b*x34[i].b;
\end{verbatim}
\item You should use proper indentation of your lines. 
This helps a great deal in recognizing
  the structure of a program. Many bugs are caused by misaligned
  braces forming a block of code. Furthermore, you should place at
  most one command per line of code. The reader will probably agree
  that 
\begin{verbatim}
for(i=0; i<number_nodes; i++)
{
  degree[i] = 0;
  for(j=0; j<number_nodes; j++)
     if(edge[i][j] > 0)
       degree[i]++;
}
\end{verbatim}
is much faster to understand than
\begin{verbatim}
for(i=0; i<number_nodes; i++) { degree[i] = 0; for(j=0;
j<number_nodes; j++)   if(edge[i][j] > 0)  degree[i]++; }
\end{verbatim}
\item Avoid jumping to other parts of a program via the
  ``goto'' \index{goto statement@{\bf goto} statement} command. 
This is bad style originating from programming in
  assembler or BASIC. In modern programming languages, 
for every logical programming  construct there are corresponding
commands. ``Goto'' commands make a program harder to understand and much
harder to debug if it does not work as it should.

In case you want to break out of a loop, you can use a while/until
loop with a flag that indicates if the loop is to be stopped. In C, if you
are lazy, you can use the commands {\tt{}break} or
{\tt{}continue}.
\item Do not use global variables. \index{global variable}
\index{variable!global} At first sight the use of global
  variables may seem tempting: you do not have to care about parameters
  for subroutines, everywhere the variables are accessible and
  everywhere they have the same name. Programming is done much faster.

But later on you will have a bad time: many bugs are created by
improper use of global variables. When you want to check for a
definition of a variable you have to search the whole list of global
variables, instead of just checking the parameter list.
Sometimes the range of validity of a global variable is
overwritten by a local variable. Furthermore, software re-usage is almost
impossible with global variables, because you always have to check
{\em all}\/ variables used in a module for conflicts 
and you are not allowed to employ
the name for another object. When you want to pass an object to a
subroutine via a global variable, you do not have the choice of how to
name the object which is to be passed. Most important, 
when you have a look onto a
subroutine after some months, you cannot see immediately which objects are
changed in the subroutine, instead 
you will have to read the whole subroutine again. 
If you avoid this practice, you just have to look at the
parameter list.
Finally, when a renaming occurs,
you have to change the name of a global variable everywhere in the
whole program. Local variables can be changed with little effort.

\item Finally, an issue of utmost importance: Do not be
  economical with comments \index{comment|(ii} in your source code! 
Most programs, which
  may appear logically structured when writing them, will be a source of
  great confusion when being read some weeks later. Every minute you
  spend on writing reasonable comments you will save later
  on several times over. You should consider different types of comments.
\begin{itemize}
\item Module comments: \index{module!comment}
At the beginning of each module you should state
  its name, what the module does, who wrote it 
and when it was written. It is a useful
  practice to include a version history, which lists the changes
  that have been performed. A module comment might look like this:
\begin{verbatim}
**********************************************************/
/*** Functions for spin glasses.                       ***/
/*** 1. loading and saving of configurations           ***/
/*** 2. initialization                                 ***/
/*** 3. evaluation functions                           ***/
/***                                                   ***/
/*** A.K. Hartmann January 1996                        ***/
/*** Version 7.0    03.07.2000                         ***/
/***                                                   ***/
/*********************************************************/

/*** Vers. History:                                    ***/
/*** 1.0 feof-check in lsg_load...() included 02.03.96 ***/
/*** 2.0 comment for cs2html added            12.05.96 ***/
/*** 3.0 lsg_load_bond_n() added              03.03.97 ***/
/*** 4.0 lsg_invert_plane() added             12.08.98 ***/
/*** 5.0 lsg_write_gen() added                15.09.98 ***/
/*** 6.0 lsg_energy_B_hom() added             20.11.98 ***/
/*** 7.0 lsg_frac_frust() added               03.07.00 ***/
\end{verbatim} 

\item Type comments: For each data type (a {\tt struct} in C or
  {\em class}\/ in C++) which you
  define in a header file, you should attach several lines of comments
describing the data type's structure and its application.  For a
class definition, also the methods  which are available should be described.
Furthermore, for a structure, each element should be explained. A nice
arrangement of the comments makes everything more readable.
An example of what such a comment may look like can be seen in
Sec.\ \ref{sec-oo} for the data type {\tt{}histo\_t}.
\item Subroutine comments: For each subroutine, its purpose, the
meaning of the input and output variables and the preconditions which
have to be fulfilled before calling must be stated. In case you are
lazy and do not write a {\em man}\/ page, a comment atop of a subroutine is the
only source of information, should you want to use the subroutine
later on in another program.

If you use some special mathematical methods or clever algorithms in
the subroutine, you should always cite the source in the comment. This
facilitates later on the understanding of how the methods works.

The next example shows what the comment for a subroutine may look like:
\begin{verbatim}
/************************* mf_dinic1() *****************/
/** Calculated maximum flow using Dinics algorithm    **/
/** See: R.E.Tarjan, Data Structures and Network      **/
/** Algorithms, p.104f.                               **/
/**                                                   **/
/** PARAMETERS: (*)= return-parameter/altered var's   **/
/**        N: number of inner nodes (without s,t)     **/
/**      dim: dimension of lattice                    **/
/**     next: gives neighbors next[0..N][0..2*dim+1]  **/
/**        c: capacities  c[0..N][0..2*dim+1]         **/
/**   (*)  f: flow values f[0..N][0..2*dim+1]         **/
/** use_flow: 0-> flow set to zero before used.       **/
/**                                                   **/
/** RETURNS:                                          **/
/**          0 -> OK                                  **/
/*******************************************************/
int mf_dinic1(int N, int dim, int *next, int *c, 
              int *f, int use_flow)
\end{verbatim}
\item Block comments: You should divide each subroutine, unless it is
  very short, into several logical blocks. A rule of thumb is that no
  block should be longer than the number of lines you can display in
  your editor window. Within one or two lines you should explain what
  is done in the block. Example:
\begin{verbatim}
  /* go through all nodes except source s and sink t in  */
  /* reversed topological order and set capacities       */
  for(t2=num_nodes-2; t2>0; t2--)
     ...
\end{verbatim}
\item Line comments: They are the lowest level 
  comments. Since you are using (hopefully) sound names for data types,
  variables and subroutines, many lines should be self
  explanatory. But in case the meaning is not obvious, you should add a small
  comment at the end of a line, for example:
\begin{verbatim}
    C(t, SOURCE) = cap_s2t[t];     /* restore capacities */
\end{verbatim}
Aligning all comments to the right makes a code easier to
read. Please avoid unnecessary comments like
\begin{verbatim}
  counter++;                         /* increase counter */
\end{verbatim}
or unintelligible comments like
\begin{verbatim}
  minimize_energy(spin, N, next, 5);   /* I try this one */
\end{verbatim}
\end{itemize}
\end{itemize}

The line containing {\tt C(t, SOURCE)} is an example of the application
of a macro. This subject is covered in the following section.
\index{programming!style|)} \index{style|)}
\index{comment|)}

\section{Programming Tools}

Programming languages and UNIX/Linux \index{UNIX} \index{Linux}
offer many  concepts and tools  
which help you to perform large
simulation projects. Here, three of them are presented: macros, 
which are explained first, {\em makefiles}\/ and scripts.

\subsection{Using Macros}
\label{sec-macros}
\index{macro|(ii}

Macros are shortcuts for code sequences in programming
languages. Their primary purpose is to
allow computer programs to be written more quickly. But the main benefit
comes from the fact that a more flexible software development
becomes possible. By using macros appropriately, programs become
better structured, more generally applicable and less error-prone.
Here it is explained how macros are defined and used in C, a detailed
introduction can be found in C textbooks such as
Ref. \cite{PRA-kernighan1988}. Other high-level programming languages
exhibit similar features.

In C a macro is  constructed via the {\tt \#define}
directive. Macros are
processed in the preprocessing stage of the compiler. This directive
has the form 
\begin{quote}
{\tt \#define}\quad {\it name} \quad {\it definition}
\end{quote}
\index{define@{\tt\#define}}

Each definition must be on one line, without other definitions or
directives. If the definition extends over more than one line, each
line except the last one has to be ended with the backslash
$\backslash$ symbol.
The simplest form of a macro is a constant, e.g.

\begin{verbatim}
#define  PI  3.1415926536
\end{verbatim}

You can use the same sorts of names for macros as for variables. It
is convention to use only upper-case letters for macros. A macro can
be deleted via the {\tt \#undef} directive.

\begin{sloppypar}
When scanning the code, the preprocessor just replaces literally 
every occurrence of a macro by its definition. If you have for example
the expression {\tt 2.0*PI*omega} in your code, the preprocessor will
convert it into {\tt 2.0*3.1415926536*omega}. You can use macros also in
the definition of other macros. But macros are not replaced in
strings, i.e.\ {\tt printf("PI");} will print {\tt PI} and not
3.1415926536 when the program is running. 
\end{sloppypar}

It is possible to test for the (non)existence of macros using the {\tt
  \#ifdef} \index{ifdef@{\tt\#ifdef}} and {\tt \#ifndef}
\index{ifndef@{\tt\#ifndef}} directives. This allows for conditional
compiling or for  platform-independent code, such as e.g.\ in

\begin{verbatim}
#ifdef UNIX
  ...
#endif
#ifdef MSDOS
  ...
#endif
\end{verbatim}

Please note that it is possible to supply definitions of macros to the
compiler via the {\tt-D} option, \index{Doption@{\tt-D} option}
e.g.\ {\tt gcc -o program program.c
-DUNIX=1}. If a macro is used only for conditional {\#ifdef/\#ifndef}
statements, an assignment like {\tt=1} can be omitted, i.e.\ {\tt-DUNIX}
is sufficient.

When programs are divided into several modules, or when library
functions are used, the definition of data types and functions are
provided in header files \index{header file} ({\tt.h} files). Each header file
should be read by the compiler only once. When projects become more
complex, many header files have to be managed, and it may become
difficult to avoid multiple scanning of some header files. This can
be prevented automatically by this simple construction using macros:

\begin{verbatim}
/** example .h file: myfile.h  **/

#ifndef _MYFILE_H_
#define _MYFILE_H_

   .... (rest of .h file)
   (may contain other #include directives)

#endif /* _MYFILE_H_ */
\end{verbatim} 

After the body of the header file has been read the first time during
a compilation process, the macro {\verb! _MYFILE_H_!} is defined, thus
the body will never read be again.

So far, macros are just constants. You will benefit from their full
power when using macros with arguments. They are given in braces
after the name of the macro, such as e.g.\ in

\begin{verbatim}
#define  MIN(x,y)  ( (x)<(y) ? (x):(y) )
\end{verbatim}

You do not have to worry more than usual 
about the names you choose for the arguments,
there cannot be a conflict with other variables of the same name,
because they are replaced by the expression you provide when a macro
is used, e.g.\ {\tt MIN(4*a, b-32)} will be expanded to {\tt
  (4*a)<(b-32) ? (4*a):(b-32)}.

The arguments are used in braces () in the macro, because the
comparison $<$ must have the lowest priority, regardless which operators
are included in the expressions that are supplied as actual arguments.
Furthermore, 
you should take care of unexpected side effects. Macros do not behave
like functions. For example when calling {\tt MIN(a++,b++)} the variable {\tt
  a} or {\tt b} may be increased twice when the program is
executed. Usually it is better to use inline functions (or sometimes
templates in C++) in such cases. But there are many applications of
macros, which cannot be replaced by incline functions, like in the
following example, which closes this section.

\begin{figure}[ht]
\begin{center}
\scalebox{0.6}{\includegraphics{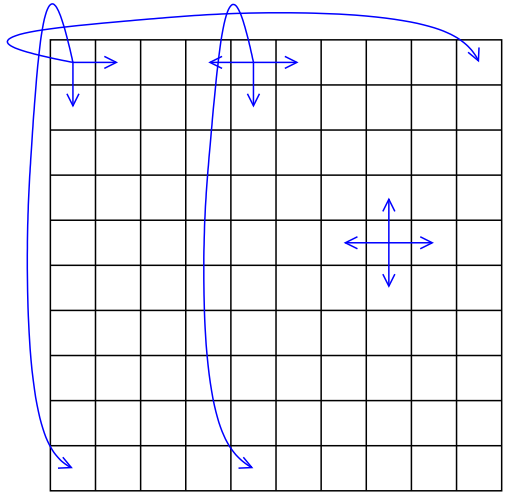}}
\caption{A square lattice of size $10\times 10$ with periodical
  boundary conditions. The arrows indicate the neighbors of the spins.}
\label{fig:pbc}
\end{center}
\end{figure}

The example illustrates how a
program can be written in a clear way using macros, making the
program less error-prone, and furthermore allowing for a broad
applicability. A system of Ising spins is considered, that is a lattice
where at each site $i$ a particle $\sigma_i$ is placed. Each particle
can have only two states $\sigma_i=\pm 1$. It is assumed  that all lattice
sites are numbered from $1$ to $N$. This is different from C arrays,
which start at index 0, the benefit of starting with index 1 for the
sites will become clear below. For the simplest version of
the model only neighbors of spins are interacting. With a
two-dimensional square lattice \index{square lattice|(}
\index{lattice!square|(} of size $N=L\times L$ a spin $i$, which is
not at the boundary, interacts with spins $i+1$ ($+x$-direction), 
$i-1$ ($-x$-direction), $i+L$ ($+y$-direction) and
$i-L$ ($-y$-direction).  A spin at the boundary may interact with fewer
neighbors when free boundary conditions are assumed. With periodic boundary
conditions (pbc), all spins have exactly 4 neighbors. In this case, 
a spin at the
boundary interacts also  with the nearest mirror images, i.e.\ with the
sites that are neighbors if you consider the system repeated in each
direction.  For a $10\times
10$ system spin 5, which is in the first row, interacts with spins
$5+1=6$, $5-1=4$, $5+10=15$ and through the pbc with spin $95$, see
Fig.\ \ref{fig:pbc}. The
spin in the upper left corner, spin 1, interacts with spins $2,11,10$
and 91. In a program pbc can be realized by performing all
calculations {\em modulo} $L$ (for the $\pm x$-directions) and 
modulo $L^2$ (for the $\pm y$-directions), respectively.

This way of realizing the neighbor relations in a program has several
disadvantages: 
\begin{itemize}
\item You have to write the code everywhere where the neighbor
  relation is needed. This makes the source code larger and less clear.
\item When switching to free boundary conditions, you have to include
  further code to check whether a spin is at the boundary.
\item Your code works only for one lattice type. If you want to extend
  the program to lattices of higher dimension you have to rewrite the
  code or provide extra tests/calculations.
\item Even more complicated would be an extension to different lattice
  structures such as triangle or face-center cubic. This would make the
  program look even more confusing. 
\end{itemize}

An alternative is to write the program directly in a way it can cope
with almost arbitrary lattice types. This can be achieved by
setting up the neighbor relation in one special initialization 
subroutine (not discussed here) and
storing it in an array {\tt next[]}. Then, the code outside the
subroutine remains the same for all lattice types and dimensions.
Since the code should work for all
possible lattice dimensions, the array {\tt next} is one
dimensional. It is assumed that each site has {\tt num\_n}
neighbors. Then the neighbors of site {\tt i} can be stored in {\tt
next[i*num\_n], next[i*num\_n+1]}, $\ldots$, {\tt next[i*num\_n+num\_n-1]}. 
Please note that the sites are numbered beginning with 1. This means,
a system with {\tt N} spins needs an array {\tt NEXT} of size {\tt
  (N+1)*num\_n}. When using free boundary conditions, missing neighbors
can be set to 0. The
access to the array can be made easier using a macro {\tt NEXT}:

\begin{verbatim}
#define  NEXT(i,r)  next[(i)*num_n + r]
\end{verbatim}

{\tt NEXT(i,r)} contains the neighbor of spin {\tt i} in direction
{\tt r}. For e.g.\ a quadratic system, {\tt r=0} is the 
$+x$-direction, {\tt r=1} the $-x$-direction, {\tt r=2} the $+y$-direction
and {\tt r=3} the $-y$-direction. However, which convention you use depends
on you, but you should make sure you are consistent. For the case of a
quadratic lattice, it is {\tt num\_n=4}. Please note that whenever the macro
{\tt NEXT} is used, there must be a variable {\verb! num_n!} defined,
which stores the number of neighbors. You could include {\verb! num_n!}
 as a third parameter of the macro, but in this case a call of
the macro looks slightly more confusing. Nevertheless, the way you
define such a macro depends on your personal preferences.

Please note that the {\tt NEXT} macro cannot be realized by an inline
function, in case you want to set values directly like in {\tt
  NEXT(i,0)=i+1}. Also, when using an inline function, you would have
to include all parameters explicitly, i.e.\ {\verb! num_n!} in the
example. The last requirement could be circumvented by using global
variables, but this is bad programming style as well.

When the system is an Ising  spin glass, the sign and magnitude of the
interaction may be different for each pair of spins. The interaction
strengths  can be
stored in a similar way to the neighbor relation, e.g.\ in an array {\tt
  j[]}.  The
access can be simplified via the macro $J$:

\begin{verbatim}
#define  J(i,r)  j[(i)*num_n + r] 
\end{verbatim}

A subroutine for calculating the energy $H=\sum_{\langle i,j \rangle}
J_{ij} \sigma_i \sigma_j$ may look as follows, please note that the
parameter {\tt N} denotes the number of spins and the values of the
spins are stored in the array {\tt sigma[]}:

\begin{verbatim}
double spinglass_energy(int N, int num_n, int *next, int *j, 
                        short int *sigma)
{
  double energy = 0.0;
  int i, r;                                           /* counters */

  for(i=1; i<=N; i++)              /* loop over all lattice sites */
    for(r=0; r<num_n; r++)             /* loop over all neighbors */
      energy += J(i,r)*sigma[i]*sigma[NEXT(i,r)];

  return(energy/2);    /* each pair has appeared twice in the sum */
}
\end{verbatim}
For this piece of code the comments explaining the parameters
and the purpose of the code are just missing for convenience. In the
actual program it should be included. 

The code for {\tt spinglass\_energy()} is very short and clear. It works
for all kinds of lattices. Only the subroutine
 where the array {\tt next[]} is set
up has to be rewritten when implementing a different type of lattice.
This is true for all kinds of code realizing e.g.\ a Monte Carlo scheme
or the calculation of a physical quantity.
For free boundary conditions, additionally {\tt sigma[0]=0} must
be assigned to be consistent with the convention that missing neighbors have
the id 0. This is the reason, why the spin site
numbering starts with index 1 while C arrays start with index 0.
\index{macro|)}
\index{square lattice|)}
\index{lattice!square|)}

\subsection{{\em Make}\/ Files}
\label{sec-make}
\index{make@{\em{}make}|(}
\index{makefile@{\em{}makefile}|(}

If your software project grows larger, it will consist of several
source-code files. Usually, there are many dependencies between the
different files, e.g.\ a data type defined in one header file can be used
in several modules. Consequently, when changing one of your source
files, it may be necessary to recompile several parts of the
program. In case you do not want to recompile your files every time by
hand, you can transfer this task to the {\em make}\/ tool which can be
found on UNIX operating systems.
A complete description of the abilities of {\em make}\/ can be found in
Ref. \cite{PRA-oram1991}. You should look on the {\em man}\/
 page \index{man page@{\em{}man} page}
(type {\tt{}man make}) or in the texinfo file \cite{PRA-texinfo}
as well. For other operating systems or
software development environments, similar tools exists. Please
consult the manuals in  case you are not working with a UNIX type of
operating system.

The basic idea of {\em make}\/ is that you keep a file which contains
all dependencies between your source code files. Furthermore, it
contains commands (e.g.\ the compiler command) which generate the
resulting files called {\em targets}\/, \index{target}
i.e.\ the final program and/or object ({\tt .o}) files.
Each pair of dependencies and commands is called {\em rule}\/.
The file containing all rules of a project is called {\em makefile}\/, 
usually it is named {\tt
  Makefile} and should be placed in the directory where the source
files are stored. 

A rule can be coded by two lines of the form

\begin{quote}
{\it target : sources}

$<$tab$>$   {\it command(s)}
\end{quote}

The first line contains the dependencies, \index{dependency} the second one the
commands. The command line must begin with a tabulator symbol {\verb!<tab>!}.
It is allowed to have several targets depending on the same sources.
You can extend the lines with the backslash ``$\backslash$'' 
at the end of each line. 
The command line is allowed to be left empty.
An example of a dependency/command pair is

\begin{verbatim}
simulation.o: simulation.c simulation.h
<tab>   cc -c simulation.c
\end{verbatim}

This means that the file {\tt simulation.o} has to be compiled if either
{\tt simulation.c} or {\tt simulation.h} have been changed. 
The {\em make}\/ program
is called by typing {\tt make} on the command line of a UNIX shell. It
uses the date of the last changes, which is stored along with each
file, to determine whether a rebuild of some targets is
necessary. Each time at least one of the  source files are newer than 
the corresponding target files, the commands given after the $<$tab$>$ are
called. Specifically, the command is called, 
if the target file does not exist at all. In this special case, no
source files have to be given after the colon in the first line of the rule.

It is also possible to generate meta rules, \index{meta rule}
which e.g.\ tell how to treat
all files which have a specific suffix. Standard rules, how to treat
 files ending for example with {\tt .c} are already included, but can be
changed for each file by stating a different rule. This subject is
covered in the {\em man}\/ page of {\em make}\/.

The make tool always tries to build only the first object of your {\em
  makefile}\/, unless enforced by the dependencies. Hence, 
if you have to build several independent object files {\tt
  object1, object2, object3}, the whole compiling must be toggled by the
first rule, thus your {\em makefile}\/ should read like this

\begin{verbatim}
all: object1 object2 object3

object1:  <sources of object1>
<tab>    <command to generate object1>

object2: ...
<tab>    <command to generate object2>

object3 ...
<tab>    <command to generate object3>
\end{verbatim}

It is not necessary to separate different rules by blank
lines. Here it is just for better readability.
If you want to rebuild just e.g.\ {\tt object3}, you can call
{\tt make object3}. This allows  several independent
targets to be combined into one {\em makefile}\/. When compiling 
programs via {\tt make}, 
it is common to include the target ``clean'' \index{clean} in the {\em makefile}\/ such that
all objects files are removed when {\tt make clean} is called. Thus,
the next call of {\tt make} (without further arguments) compiles the
whole program again from scratch. The rule for `clean` reads like

\begin{verbatim}
clean:
<tab>   rm -f *.o
\end{verbatim}

Also iterated dependencies are allowed, for example

\begin{verbatim}
object1: object2 

object2: object3
<tab> ...

object3: ...
<tab> ...
\end{verbatim}

The order of the rules is not important, except that {\em make}\/
always starts with the first target.
Please note that the {\em make}\/ tool is not just
intended to manage the software development process 
and toggle compile commands. Any
project where some output files depend on some input files in an
arbitrary way can be controlled. For example you could control the
setting of a book, where you have text-files, figures, a bibliography
and an index as input files. The different chapters and finally the
whole book are the target files.

Furthermore, it is possible to define variables, sometimes also called
macros. They have the format

\begin{quote} 
{\it variable}={\it definition}
\end{quote}

Also variables belonging to your environment like {\tt \${}HOME} can be
referenced in the {\em makefile}\/.
The value of a variable can be used, similar to shells variables, 
by placing a \$ sign in front of the name of the variable, but you
have to embrace the name by $(\ldots)$ or
$\{\ldots\}$. There are some special variables, e.g.\ \$@ holds the
name of the target in each corresponding command line, here no braces
are necessary.
The variable {\tt CC} is predefined to hold the compiling command, you
can change it by including for example

\begin{verbatim}
CC=gcc
\end{verbatim}

in the {\em makefile}\/.
In the command part of a rule the compiler is called via {\tt
  \${}(CC)}. Thus, you can change your compiler for the whole project
very quickly by  altering just one line of the {\em makefile}\/. 

Finally, it will be shown what a typical {\em makefile}\/ for a small software
project might look like. The resulting program is called 
{\tt{}simulation}. There are two additional modules {\tt{}init.c},
{\tt{}run.c} and the corresponding header
{\tt.h} files. In {\tt{}datatypes.h} types are defined which are
used in all modules. Additionally, an external precompiled object file
{\tt{}analysis.o} in the directory {\tt\${}HOME/lib} is to be linked, the
corresponding header file is assumed to be stored in
{\tt\${}HOME/include}. For {\tt{}init.o} and {\tt{}run.o} no
commands are given. In this case {\tt make} applies the predefined
standard command for files having {\tt.o} as suffix, which reads like

\begin{verbatim}
<tab>   $(CC) $(CFLAGS) -c $@
\end{verbatim}

where the variable {\tt{}CFLAGS} may contain options passed to the
compiler and is initially empty.
The {\em makefile}\/ looks like this, please note that
lines beginning with ``\#''  are comments.

\begin{verbatim}
#
# sample make file
#
OBJECTS=simulation.o init.o run.o 
OBJECTSEXT=$(HOME)/lib/analysis.o
CC=gcc
CFLAGS=-g -Wall -I$(HOME)/include
LIBS=-lm 

simulation: $(OBJECTS) $(OBJECTSEXT)
<tab>   $(CC) $(CFLAGS) -o $@ $(OBJECTS) $(OBJECTSEXT) $(LIBS)

$(OBJECTS): datatypes.h

clean:
<tab>   rm -f *.o
\end{verbatim}

The first three lines are comments, then five variables {\tt{}OBJECTS},
{\tt{}OBJECTSEXT}, {\tt{CC}}, {\tt{}CFLAGS} and {\tt{}LIBS} are
assigned. The final part of the {\em makefile}\/ are the rules.

Please note that sometimes bugs are introduced, if the {\em makefile}\/
is incomplete. For example consider a header file which is included in
several code files, but this is not mentioned in the {\em
  makefile}\/. Then, if you change e.g.\ a data type in the header file,
some  of the code files might not be compiled again, especially
those you did not change. Thus the same objects files can be treated with
different formats in your program, yielding bugs which seem hard to
explain. Hence, in case you encounter mysterious bugs, a {\tt make
  clean} might help. But most of the time, bugs which are hard to
explain are due to errors in your memory management. How to track
down those bugs  is explained in Sec.\ \ref{sec-testing}.

The {\tt{}make} tool exhibits many other features. For additional details, 
please consult the references given above.
\index{make@{\em{}make}|)}
\index{makefile@{\em{}makefile}|)}

\subsection{Scripts}
\index{script|(ii}

Scripts are even more general tools than {\em make}\/ files. 
They are in fact small programs,
but they are usually not compiled, i.e.\ they are quickly written but
they run slowly. Scripts can be used to perform many administration tasks
like backing up data, installing software or running simulation
programs for many different parameters. 
Here only an example concerning the last task is
presented. For a general introduction to scripts, please refer to a
book on UNIX/Linux.

Assume that you have a simulation program called {\tt coversim21}
which calculates vertex covers of graphs. In case you do not know what
a vertex cover is, it does not matter, just regard it as one
optimization problem characterized by some parameters.
You want to run the program
for a fixed graph size {\tt L}, for a fixed concentration {\tt c} of
the edges, average over {\tt num} realizations and write the results
to a file, which contains a string {\tt appendix} in its name to
  distinguish it from other output files. Furthermore, you want to
  iterate over different relative sizes {\tt x}. Then you can use the following
  script {\tt run.scr}:

\begin{verbatim}
#!/bin/bash
L=$1
c=$2
num=$3
appendix=$4
shift
shift
shift
shift
for x
do
  ${HOME}/cover/coversim21 -mag  $L $c $x $num > \
                       mag_${c}_${x}${appendix}.out
done
\end{verbatim}
The first line starting with ``\#'' is a comment \index{comment}
line, but it has a
special meaning. It tells the operating system the language in which the
script is written. In this case it is for the  {\em bash}\/ shell, the
absolute pathname of the shell is given. Each
UNIX shell has its own script language, you can use all commands
which are allowed in the shell. There are also more elaborate script
languages like {\em perl}\/ \index{perl@{\em{}perl}} or 
{\em phyton}\/, \index{phyton@{\em{}phyton}} but they are not covered here.

Scripts can have command line arguments, which are referred via 
{\verb!$1!}, {\verb!$2!}, {\verb!$2!}
etc., the name of the script itself is stored in {\verb!$0!}. Thus, in
the lines 2 to 5, four variables are assigned. In 
general, you can use the arguments everywhere in the script directly,
i.e.\ it is not necessary to store them in other variables. It is done
here because in the next four lines the arguments {\verb!$1!} to
{\verb!$4!} are
thrown away by four {\tt shift} commands. Then, the argument which was
on position five at the beginning is stored in the first
argument. Argument zero, containing the script name, is not affected
by the shift.

Next, the script enters a loop, given by ``{\tt{}for x; do
  ... done}''. This construction means that iteratively all remaining
arguments are assigned to the variable ``{\tt{}x}'' and each time the body
of the loop is executed. In this case, the simulation is started
with some parameters and the output directed to a file. Please note
that you can state the loop parameters explicitly like in ``{\tt{}for
  size in 10 20 40 80 160; do ... done}''.

The above script can be called for example by 

\begin{verbatim}
run.scr 100 0.5 1000 testA 0.20 0.22 0.24 0.26 0.28 0.30
\end{verbatim}

which means that the graph size is 100, the fraction of edges is 0.5, the
number of realizations per run is 100, the string {\tt testA}
appears in the output file name and the simulation is performed for 
the relative sizes 0.20, 0.22, 0.24, 0.26, 0.28, 0.30.
\index{script|)}

\section{Libraries}
\label{sec-lib}
\index{library|(ii}

Libraries are collections of subroutines and data types, which can be
used in other programs. There are libraries for numerical methods
such as integration or solving differential equations, for storing,
sorting and accessing data, for fancy data types like lists or trees,
for generating colorful graphics and for thousands of other
applications. Some can be obtained for free, while other, usually
specialized libraries have to be purchased. The use of libraries
speeds up the software development process enormously, because you
do not have to implement every standard method by yourself. Hence, you
should always check whether someone has done the jobs for you already,
before starting to write a program. Here, two standard libraries are briefly
presented, providing routines which are needed for
most computer simulations. 

Nevertheless, sometimes it is inevitable to implement some
methods by yourself. 
In this case, after the code has been proven to be reliable
and useful for some time, you can put it in a self-created
library. How to create libraries is explained in the last part of
this section.

\subsection{Numerical Recipes}
\index{Numerical Recipes|(ii}

The {\em Numerical Recipes}\/ (NR)
\cite{PRA-numrec1995} contain a huge number of subroutines
to solve standard numerical problems. Among them are:
\begin{itemize}
\item solving linear equations \index{linear equation}
\item performing interpolations \index{interpolation}
\item evaluation and integration of functions \index{integration}
\item solving nonlinear equations \index{nonlinear equation}
\item  minimizing functions
\item diagonalization of matrices \index{matrix!diagonalization} 
\item Fourier transform \index{Fourier transform}
\item solving ordinary and partial differential equations.
\index{differential equation} 
\end{itemize}
The algorithms included are all
state of the art. There are several libraries dedicated to similar
problems, e.g.\ the library of the {\em Numerical Algorithms Group}\/ \cite{PRA-nag}
or the subroutines which are included with the {\em Maple}\/
\index{Maple} software package
\cite{PRA-maple}. 

To give you an impression how the subroutines can be used,
just a short example is presented. 
Consider the case that a symmetrical matrix is given and that all
eigenvalues are to be determined. 
For more information on the library the reader should consult 
Ref. \cite{PRA-numrec1995}. There it is not only shown how the
library can be applied, but also all algorithms are explained. 

The program to calculate the eigenvalues \index{eigenvalue} reads as follows.
\begin{verbatim}
#include <stdio.h>
#include <stdlib.h>
#include "nrutil.h"
#include "nr.h"

int main(int argc, char *argv[])
{
  float **m, *d, *e;                      /* matrix, two vectors */
  long n =  10;                                /* size of matrix */
  int i, j;                                      /* loop counter */

  m = matrix(1, n, 1, n);                     /* allocate matrix */
  for(i=1; i<=n; i++)              /* initialize matrix randomly */
    for(j=i; j<=n; j++)
    {
        m[i][j] = drand48();
        m[j][i] = m[i][j];      /* matrix must be symmetric here */
    }

  d = vector(1,n);        /* contains diagonal elements          */
  e = vector(1,n);        /* contains off diagonal elements      */
  tred2(m, n, d, e);      /* convert symmetric m. -> tridiagonal */
  tqli(d, e, n, m);       /* calculate eigenvalues               */
  for(j=1; j<=n; j++)     /* print result stored now in array 'd'*/
    printf("ev %d = %f\n", j, d[j]);

  free_vector(e, 1, n);                      /* give memory back */
  free_vector(d, 1, n);
  free_matrix(m, 1, n, 1, n);
  return(0);
}
\end{verbatim}

In the first part of the program, an $n\times n$ matrix is allocated via the
subroutine {\tt matrix()} which is provided by {\em Numerical Recipes}\/. It is
standard to let a vector start with index 1, while in C usually a
vector starts with index 0.

In the second part a matrix is initialized randomly. Since the
following subroutines work only for symmetric real matrices, the matrix
is initialized symmetrically. The {\em Numerical Recipes}\/ also provide
methods to diagonalize arbitrary matrices, for simplicity
this special case is chosen here .

In the third part the main work is done by the {\em  Numerical Recipes}\/
subroutines {\tt tred2()} \index{tred2@{\tt{}tred2()}}
and {\tt tqli()}. \index{tqli@{\tt{}tqli()}} First, the matrix is
written in tridiagonal form by a Householder transformation 
\index{Householder transformation}
({\tt tred2()}) and then the actual eigenvalues are calculated by
calling {\tt tqli(d, e, n, m)}. The eigenvalues are returned in the
vector {\tt d[]} and the eigenvectors in the matrix {\tt m[][]}
(not used here), which is overwritten. 
Finally the memory allocated for the matrix and the
vectors is freed again.

This small example should be sufficient to show how simply the
subroutines from the {\em Numerical Recipes}\/ can be incorporated into a
program. When you have a problem of this kind you should always
consult the NR library first, before starting to write code by yourself.
\index{Numerical Recipes|)}

\subsection{LEDA}
\index{LEDA library|(ii}

While the {\em Numerical Recipes}
 are dedicated to numerical problems, the
{\em Library of Efficient Data types and Algorithms}\/ (LEDA) \cite{PRA-leda1999}
can help a great deal in  
writing efficient programs in general. It is written in C++, but it can be
used by C style programmers as well via mixing C++ calls to LEDA
subroutines within C code. LEDA contains many basic and advanced data
types such as:
\begin{itemize}
\item strings \index{string}
\item numbers of arbitrary precision \index{arbitrary precision}
\item one- and two-dimensional arrays \index{array}
\item lists \index{list} and similar objects like stacks 
\index{stack} or queues \index{queue}
\item sets \index{set}
\item trees \index{tree}
\item graphs (directed and undirected, also labeled) \index{graph}
\item dictionaries, \index{dictionary}
there you can store objects with arbitrary key  words as indices
\item data types for two and three dimensional geometries, like
  points, segments or spheres
\end{itemize}

For most data types, it is possible to create arbitrary complex
structures by using templates. For example you can make lists of self
defined structures or stacks of trees. The most efficient
implementations known in literature so far are taken for all data
structures. Usually, you can choose between different implementations,
to match special requirements. For every data type, all
necessary operations are included; e.g.\ for lists:
\index{list!operations}  creating, appending,
splitting, printing and deleting lists as well as inserting,
searching, sorting and deleting elements in a list, also iterating over all
elements of a list. The major part of the library is dedicated to
graphs and related algorithms. You will find for example subroutines to
calculate strongly connected components, shortest paths, maximum flows,
minimum cost flows and (minimum) matchings. 
\index{graph!algorithms in LEDA}

Here again, just a
short example is given to illustrate how the library can be utilized and to
show how easy LEDA can be used. 
A list of a self defined class {\tt Mydatatype} is considered. Each
element contains the data entries {\tt info} and {\tt flag}. In the first part
of the program below, the class {\tt Mydatatype} is
partly defined. Please note that input and output stream operators
{\tt{}<<}/{\tt{}>>} must be provided to be able to create a list of
{\tt{}Mydatatype} elements, otherwise the program will not compile. 
In the main part of the program a list is
defined via the LEDA data type {\tt{}list}. Elements are inserted
into the list with {\tt{}append()}. Finally an iteration over all
list elements is performed using the LEDA macro {\tt{}forall}. The
program {\tt leda\_test.cc} reads as follows:
\begin{verbatim}
#include <iostream.h>
#include <LEDA/list.h> 

class Mydatatype                // self defined example class 
{
 public:
    int       info;                            // user data 1
    short int flag;                            // user data 2
    Mydatatype() {info=0; flag=0;};            // constructor
    ~Mydatatype() {};                          // destructor
    friend ostream& operator<<(ostream& O, const Mydatatype& dt) 
        { O << "info: " << dt.info << " flag: " << dt.flag << "\n"; 
          return(O);};                         // output operator
    friend istream& operator>>(istream &I, Mydatatype& dt) 
        {return(I);};                          // dummy
};
\end{verbatim}
\clearpage
\begin{verbatim}
int main(int argc, char *argv[])
{
  list<Mydatatype> l;       // list with elements of 'Mydatatype' 
  Mydatatype element;
  int t;

  for(t=0; t<10; t++)       // create list
  {
    element.info = t;
    element.flag = t%2;
    l.append(element);
  }
  forall(element, l)        // iterate over all elements 
    if(element.flag)        // print only 'even' elements
      cout << element;
  return(0);
}
\end{verbatim}

The program has to be compiled with a C++ compiler. Depending on your
system, you have to specify some compiler flags to include LEDA,
please consult your systems documentation or the system
administrator. The compile command may look like this:

\begin{verbatim}
g++ -I$LEDAROOT/incl -L$LEDAROOT -o leda_test leda_test.cc -lG -lL
\end{verbatim}

The {\tt -I} flag specifies where the compiler searches for header files
like {\tt LEDA/list.h}, the {\tt -L} flag tells where the libraries
({\tt -lG -lL}) are located. The environment variable {\tt LEDAROOT}
must point to the directory where LEDA is stored in your system.

Please note that using {\em Numerical Recipes}\/ and LEDA together results in
conflicts, since the objects {\tt vector} and {\tt matrix} are defined
in both libraries. You can circumvent this problem by taking the 
source code of {\em Numerical Recipes}\/ (here: {\tt nrutil.c}, {\tt
  nrutil.h}) and rename the subroutines {\tt matrix()} and {\tt
  vector()}, compile again and include {\tt nrutil.o} directly in your
program. 

Here, it should be stressed: Before trying to write everything by
yourself, you should check whether someone else has done it
for you already. LEDA is a highly effective and very convenient tool. It will
save you a lot of time and effort when you use it for your program
development. 
\index{LEDA library|)}

\subsection{Creating your own Libraries}
\index{library!create|(}

Although many useful libraries are available, sometimes you
have to write some code by yourself. Over the years you will collect
many subroutines, which -- if properly designed -- can be included in
other programs, in which  case it is convenient to put these subroutines
in a library. Then you do not have to include the object file every
time you compile one of your programs. If your self-created library is
put in a standard search path, you can access it like a system library,
you even do not have to remember where the object file is stored.

To create a library you must have an object file, e.g.\ {\tt tasks.o},
and a header file {\tt tasks.h} where all data types and function
prototypes are defined. Furthermore, to facilitate the use of the
library, you should write a {\em man}\/ page, which is not necessary for
technical reasons but results in a more convenient usage of your
library, particularly should other people want to benefit from it. 
To learn how to write a {\em man} page 
\index{man page@{\em{}man} page} you should consult
{\tt man man} and have a look at the source code of some {\em man}\/ pages,
they are stored e.g.\ in {\tt /usr/man}. 

A library is created with the UNIX command {\tt ar}. To include {\tt
  tasks.o} in your library {\tt libmy.a} you have to enter
\begin{verbatim}
ar r libmy.a tasks.o
\end{verbatim}
In a library several object files may be collected. 
The option ``{\tt r}'' replaces the given object files, if they
already belong to the library, otherwise they are added. If the library does
not exist yet it is created. For more options, please refer to the
{\em man}\/ page of {\tt ar}.

After including an object file, you have to update an internal
object table of the library. This is done by
\begin{verbatim}
ar s libmy.a
\end{verbatim}

Now you can compile a program {\tt prog.c} using your library via
\begin{verbatim}
cc -o prog prog.c libmy.a
\end{verbatim}

In case {\tt libmy.a} contains several object files, it saves some typing
by just writing {\tt libmy.a}, furthermore you do not have to remember
the names of all your object files.

To make the handling of the library more comfortable, you can create a
directory, e.g.\ {\tt $\sim$/lib} and put your libraries
there. Additionally, you should create the directory {\tt $\sim$/include}
where all personal header files can be collected. Then your compile
command may look like this:
\begin{verbatim}
cc -o prog prog.c -I$HOME/include -L$HOME/lib -lmy
\end{verbatim}
The option {\tt -I} states the search path for additional header
files, the {\tt -L} option tells the linker where your libraries are
stored and via {\tt -lmy} the library {\tt libmy.a} is
actually included. Please note that the prefix {\tt lib} and the postfix
{\tt.a} are omitted with the {\tt -l} option. Finally, it should be
pointed out, that the compiler command given above works in all
directories, once you have set up the structure as explained. Hence,  you
do not have to remember directories or names of object files.
\index{library|)}
\index{library!create|)}

\section{Random Numbers}
\label{sec-random}
\index{random number generator|(}

For many simulations in physics, random numbers are necessary. Quite
often the model itself exhibits random parameters which remain  fixed 
throughout the simulation, one speaks of {\em quenched
  disorder}\/. \index{quenched disorder} \index{disorder!quenched}
A famous example are spin glasses. \index{spin glass} In this case one has
to perform an average over different realizations of the disorder, to
obtain physical quantities.

But even when the system which is treated is not random, very often
random numbers are required by the algorithms, e.g.\ to realize a
finite-temperature ensemble or when using randomized algorithms. In
this section an introduction to the generation of random
numbers is given. First it is explained how they can be generated at all on a
computer. Then, different methods for obtaining numbers are explained, 
which obey a given distribution: the {\em inversion method}\/, the 
{\em Box-M\"uller
method}\/ and the {\em rejection method}\/. More comprehensive information about
these and similar techniques can be found in
Refs. \cite{PRA-numrec1995,PRA-morgan1984}. 

In this section it is assumed that you are familiar with the basic
concepts of probability theory and statistics.

\subsection{Generating Random Numbers}

First, it should be pointed out that standard computers are
deterministic machines. Thus, it is completely impossible  to generate
true random numbers, at least not without the help of the user. It is
for example possible to measure the time interval between successive
keystrokes, which is randomly distributed by nature. But they depend
heavily on the current user and it is not possible to reproduce an
experiment in exactly the same way. This is the reason why
{\em pseudo random numbers}\/ are usually taken. They are generated by
deterministic rules, but they look like and have many of the
properties of true random numbers. One would like to have a random
number generator {\tt{rand()}}, such that each possible number has the same
probability of occurrence. Each time {\tt{}rand()} is called, a new
random number is returned.
Additionally, if two numbers $r_i,r_k$ differ
only slightly, the random numbers $r_{i+1},r_{k+1}$ returned by the
respective subsequent calls should have a low
correlation. 

The simplest methods to generate pseudo random numbers are 
{\em linear congruential generators}\/. 
\index{linear congruential generators} They generate a sequence
$I_1,I_2,\ldots$ of integer numbers between 0 and $m-1$ by a recursive
recipe:
\begin{equation}
  I_{n+1} = (aI_n+c) \mbox{mod} m
\end{equation}
To generate random numbers $r$ distributed in the interval $[0,1)$ one
has to divide the current random number by $m$. It is desirable to
obtain equally distributed values in the interval, i.e.\ a uniform
distribution. Below, you will see,
how random numbers obeying other distributions can be generated from
uniformly distributed numbers. 

 The real art is to
choose the parameters $a,c,m$ in a way that ``good'' random numbers are
obtained, where ``good'' means ``with less correlations''. 
In the past several results from simulations have been turned out
to be wrong, because of the application of bad random number
generators \cite{PRA-randomTest}. 

\clearpage
\begin{example}{Bad and good generators}
  To see what ``bad generator'' means, consider as an example the
parameters $a=12351,c=1,m=2^{15}$ and the seed value $I_0=1000$. 
10000 random numbers are generated, by dividing each of
them by $m$, they are distributed in the interval $[0,1)$. In
Fig.\ \ref{fig:randomDistr} the distribution of the random numbers is shown.
\begin{figure}[ht]
\begin{center}
\scalebox{0.5}{\includegraphics{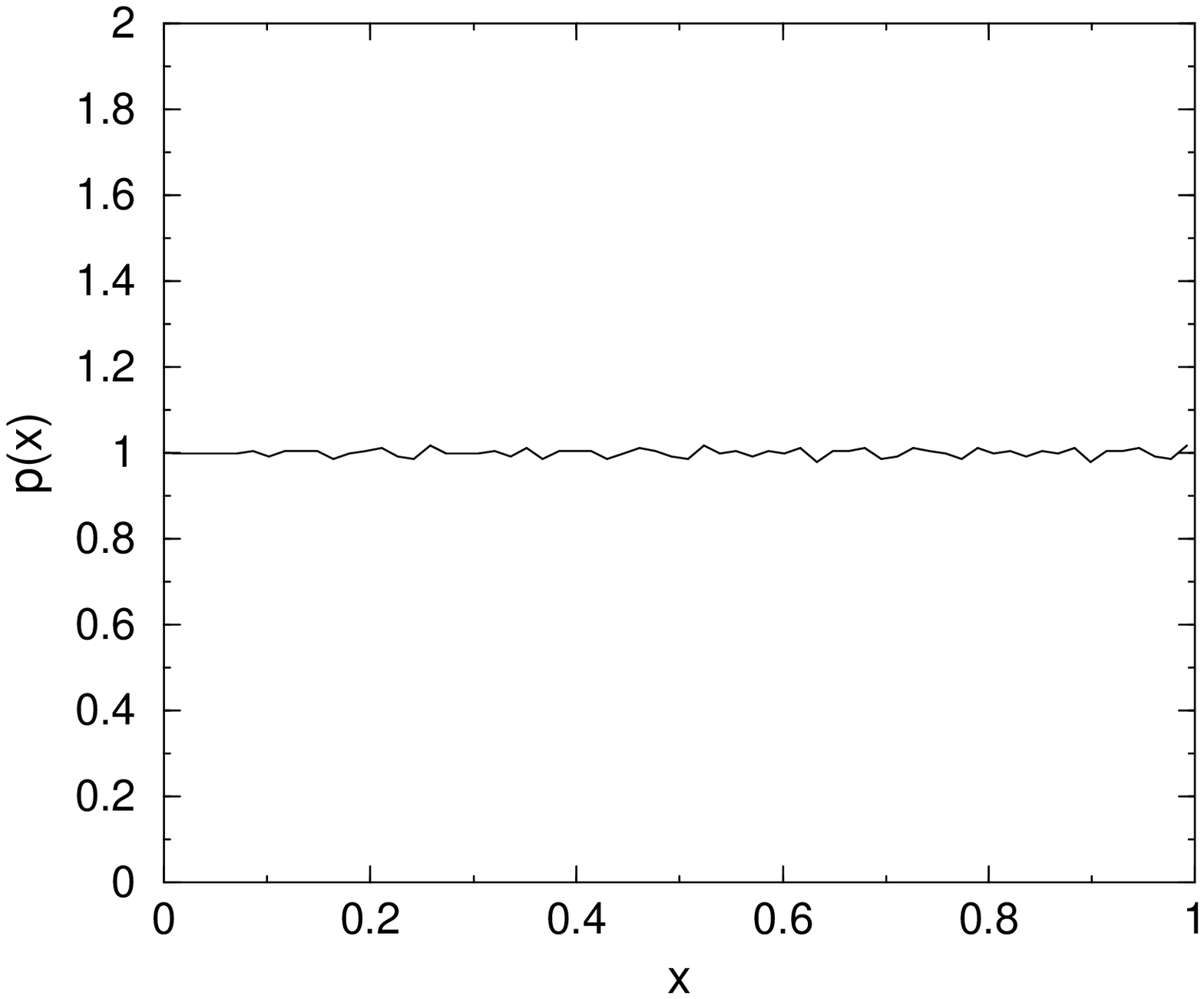}}
\narrowcaption{Distribution of random numbers in the interval $[0,1)$. They
  are generated using a linear congruential generator with the
  parameters $a=12351,c=1,m=2^{15}$.}
\label{fig:randomDistr}
\end{center}
\end{figure}

\vspace{-1mm}
\begin{sloppypar}
The distribution looks rather flat, but by taking a closer look some
regularities can be observed. These regularities can be studied
 by recording $k$-tuples of $k$ successive random numbers
$(x_i,x_{i+1},\ldots, x_{i+k-1})$. A good random number generator, 
exhibiting no correlations,
would fill up the $k$-dimensional space uniformly. 
Unfortunately, for linear congruential generators, 
instead the points lie on $(k-1)$-dimensional planes. It can be shown
that there are {\em at most}\/ of the order $m^{1/k}$ such planes. A bad
generator has much fever planes. This is the case for the example
studied above, see top part of Fig.\ \ref{fig:randomCorr}
\end{sloppypar}
\begin{figure}[!ht]
\begin{center}
\scalebox{0.5}{\includegraphics{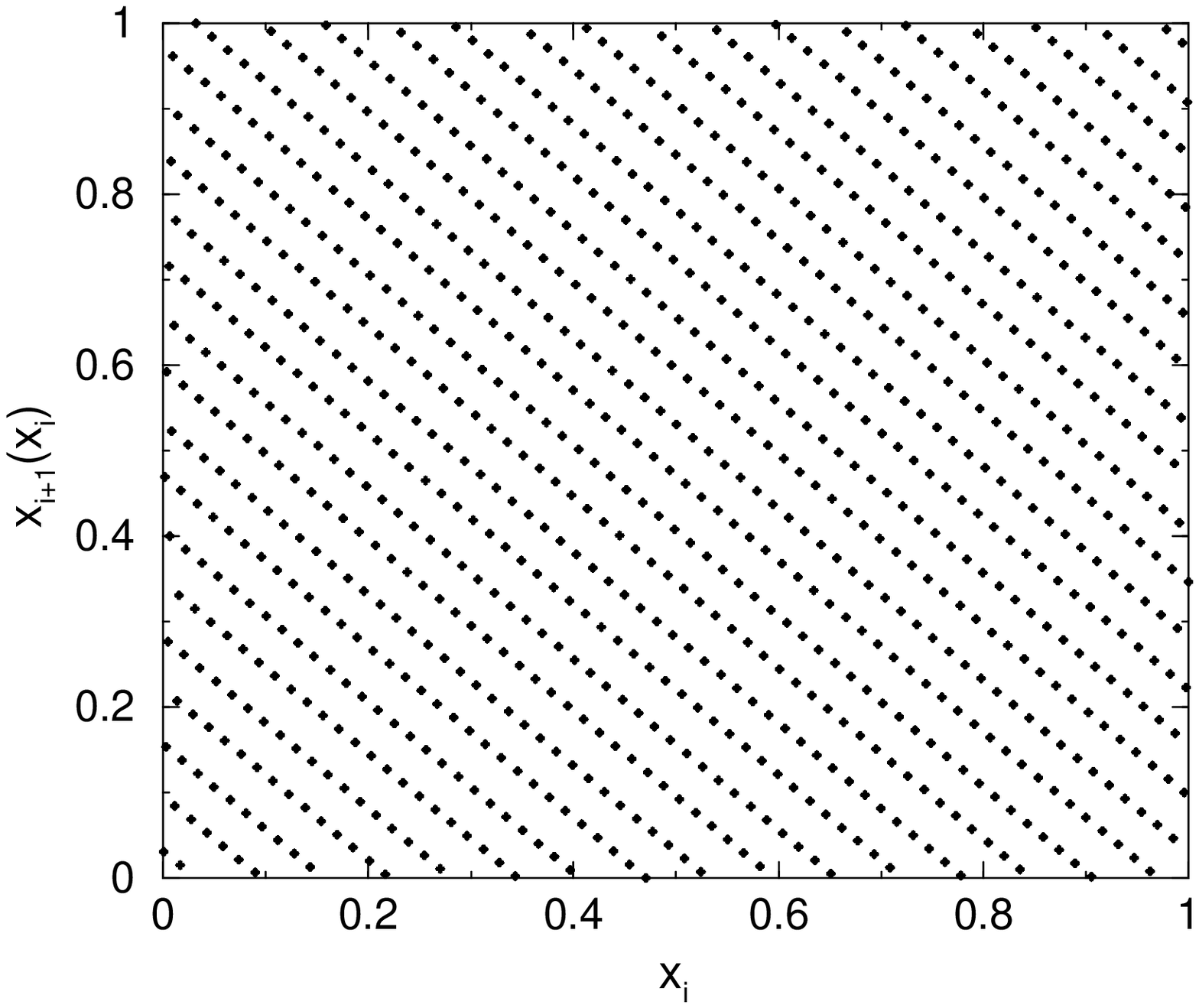}}
\scalebox{0.5}{\includegraphics{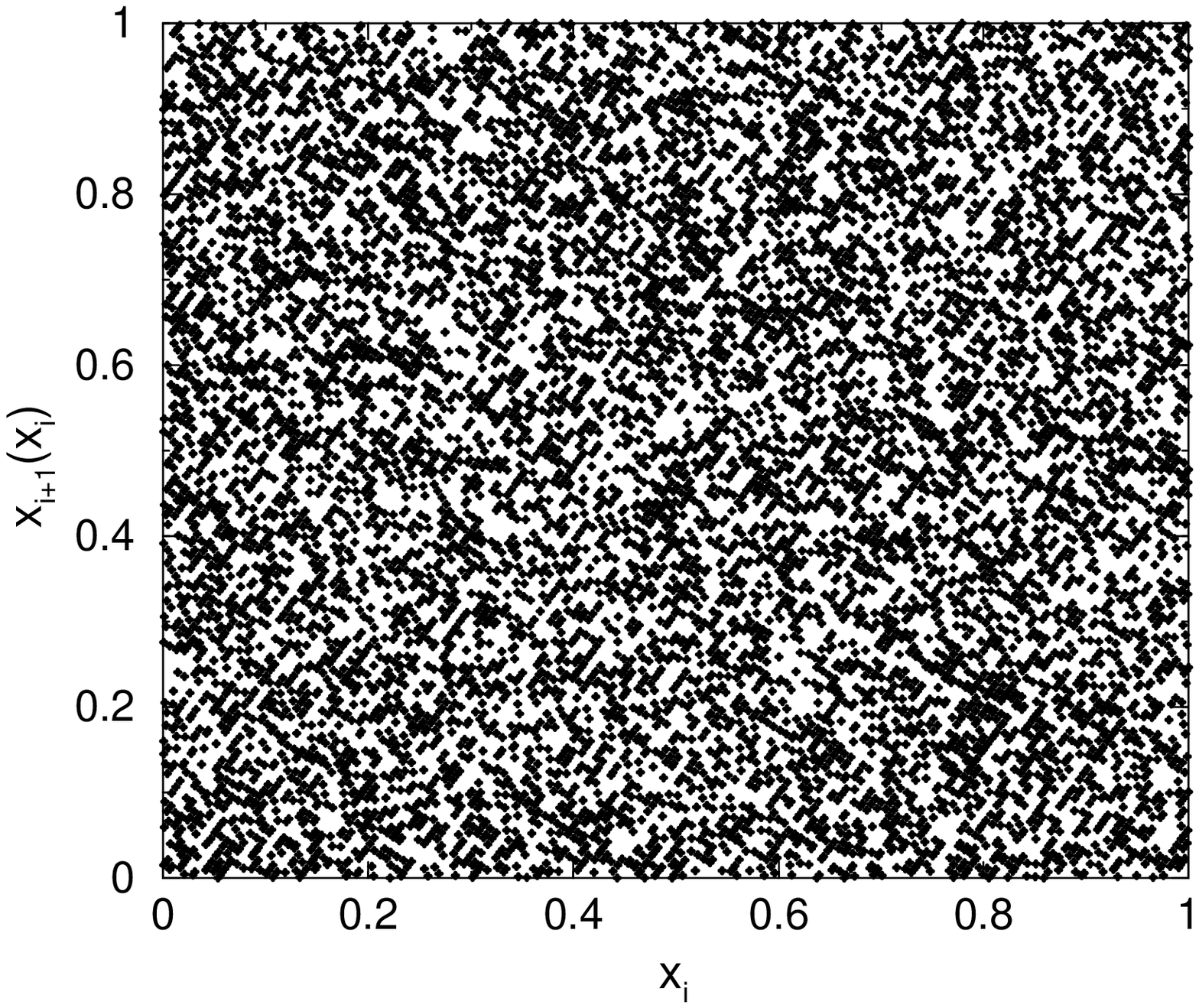}}
\narrowcaption{Two point correlations $x_{i+1}(x_i)$ 
between successive random numbers $x_i,x_{i+1}$. The top case
  is generated using a linear congruential generator with the
  parameters $a=12351,c=1,m=2^{15}$, the bottom case has instead $a=12349$.}
\label{fig:randomCorr}
\end{center}
\end{figure}

\vspace{-1.5mm}
The result for $a=123450$ is even worse, only 15 different ``random''
numbers are generated (with seed 1000), then the iteration reaches a
fixed point (not shown in a figure).

\vspace{-1.5mm}
If instead $a=12349$ is chosen, the two-point correlations look like
that shown in the bottom half of 
Fig.\ \ref{fig:randomCorr}. Obviously, the behavior is much more
irregular, but poor correlations may become visible for  higher $k$-tuples.
\end{example}
\clearpage
A generator which has passed several theoretical test is
$a=7^5=16807$, $m=2^{31}-1$, $c=0$. When implementing this 
generator you have to
be careful, because during the calculation numbers are generated which
do not fit into 32 bit. A clever implementation is presented in
Ref. \cite{PRA-numrec1995}. Finally, it should be stressed that this
generator, like all linear congruential generators, has the low-order
bits much less random than the high-order bits. For that reason, when
you want to generate integer numbers in an interval [1,N], you should
use
\begin{verbatim}
 r = 1+(int) (N*(I_n)/m);
\end{verbatim}
 instead of using the modulo operation as with {\tt r=1+(I\_n \% N);}.

So far it has been shown how random numbers can be
generated which are distributed uniformly in the interval $[0,1)$. In
general, one is interested in obtaining random numbers which are
distributed according to a given probability distribution with density
$p(z)$. In the next sections, several techniques performing this task
for continuous probability distributions are presented. 

In case of discrete distributions, one has to create
a table of the possible outcomes with their probabilities $p_i$.
To draw a number, one has to draw a  
random number $u$ which is uniformly distributed in $[0,1)$  
and take the entry $j$ of the table such that the
sum $\sum_{i=1}^j p_i$ of the preceding probabilities is larger than
$u$, but $\sum_{i=1}^{j-1}p_i<u$.
In the following, we concentrate on techniques for 
generating continuous random variables.

\subsection{Inversion Method}
\index{inversion method|(}

Given is a random number generator $drand()$ which is assumed to
generate random numbers $U$ which are distributed uniformly in $[0,1)$. The
aim is to generate random numbers $Z$ with probability density
$p(z)$. The  corresponding distribution function is

\begin{equation}
  P(z)\equiv \mbox{Prob}(Z\le z) \equiv
\int_{-\infty}^z dz^{\prime} p(z^{\prime}) \label{eq:pra:transform}
\end{equation}

The target is to find a function $g(X)$, such that after the 
transformation $Z=g(U)$,
the values of $Z$ are distributed according to
(\ref{eq:pra:transform}). It is assumed that $g$ can be inverted and is
strongly monotonically increasing, then one obtains
\begin{equation}
  P(z)=\mbox{Prob}(Z\le z)=\mbox{Prob}(g(U)\le z) = \mbox{Prob}(U\le g^{-1}(z))
\end{equation}
Since the distribution function $F(u)=\mbox{Prob}(U\le u)$ for a uniformly 
distributed variable is just $F(u)=u$ ($u\in[0,1]$), one obtains
$P(z)=g^{-1}(z)$. Thus, one just has to choose $g(z)=P^{-1}(z)$ for the
transformation function, in order to obtain random numbers, which are
distributed according the probability distribution $P(z)$. Of course,
this only works if $P$ can be inverted.

\clearpage

\begin{example}{Exponential distribution}
\index{exponential distribution} \index{distribution!exponential}
  Let us consider the exponential distribution with parameter $\lambda$,
  with probability density 
  \begin{equation}
    p(z)=\lambda\exp(-\lambda z)
  \end{equation}
and distribution function $P(z)=1-\exp(-\lambda z)$. Therefore, one can
obtain exponentially distributed random numbers $Z$, by generating
uniform distributed random numbers $U$ and choosing
$Z=-\ln(1-U)/\lambda$.

\begin{figure}[ht]
\begin{center}
\scalebox{0.5}{\includegraphics{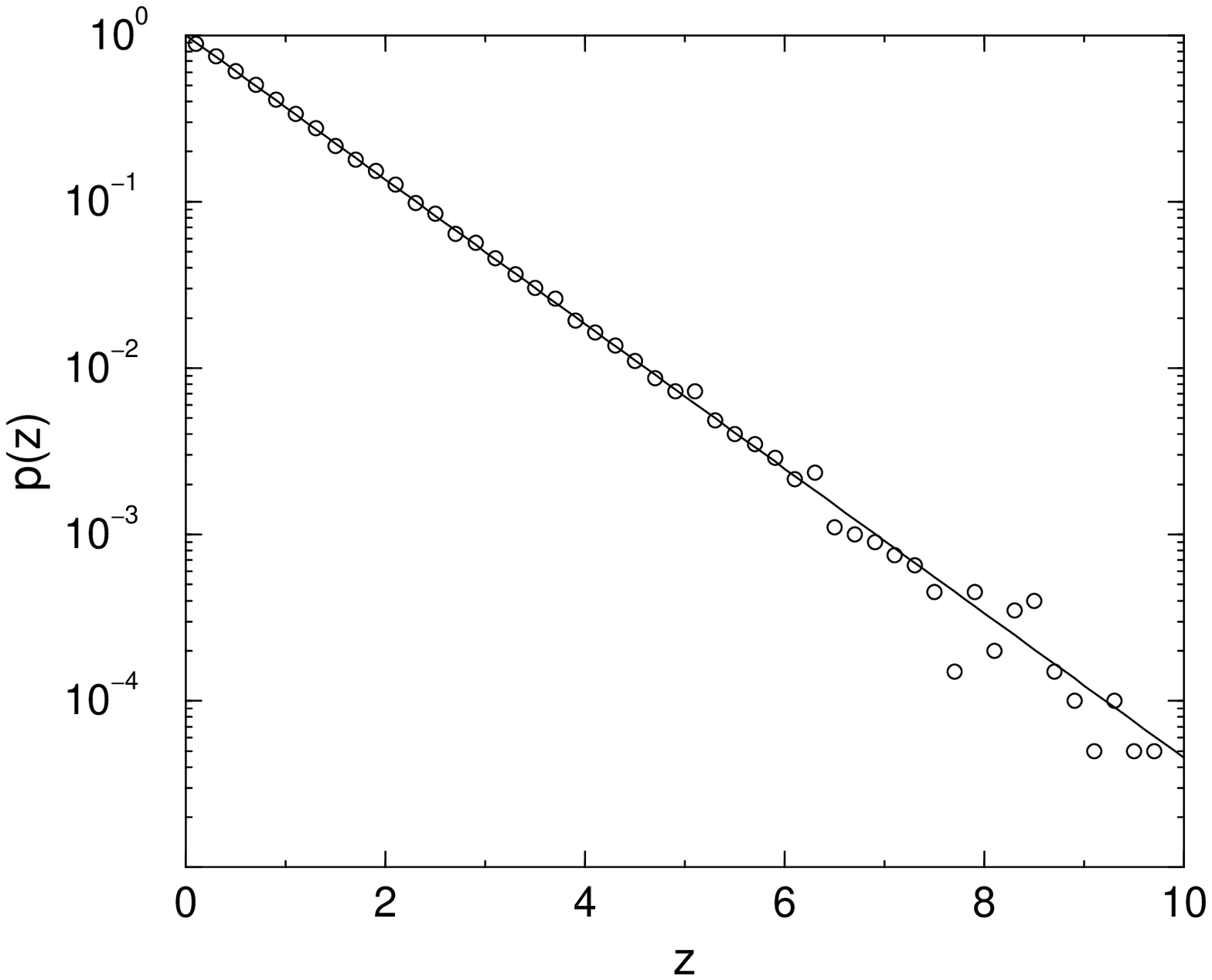}}
\narrowcaption{Histogram of  random numbers
  generated according to an exponential distribution
  ($\lambda=1$) compared with the probability density (straight line) in
  a logarithmic plot.}
\label{fig:exponential}
\end{center}
\end{figure}
In Fig.\ \ref{fig:exponential} a histogram for $10^5$ random numbers
generated in this way and the exponential probability function for
$\lambda=1$ are shown with a logarithmically scaled $y$-axis. Only for
larger values are deviations  visible. They are due to statistical
fluctuations since $p(z)$ is very small there.

For completeness, 
this example is finished by mentioning that by summing $n$ independent
exponentially distributed random numbers, the result is gamma
distributed \cite{PRA-morgan1984}.
\end{example}
\index{inversion method|)}

\subsection{Rejection Method}
\index{rejection method|(}

As mentioned above, the inversion method works only when the
distribution function $P$ can be inverted. For distributions not
fulfilling this condition, sometimes this problem can
be overcome by drawing several random numbers and combining them in a
clever way, see e.g.\ the next subsection.

\begin{figure}[ht]
\begin{center}
\scalebox{0.5}{\includegraphics{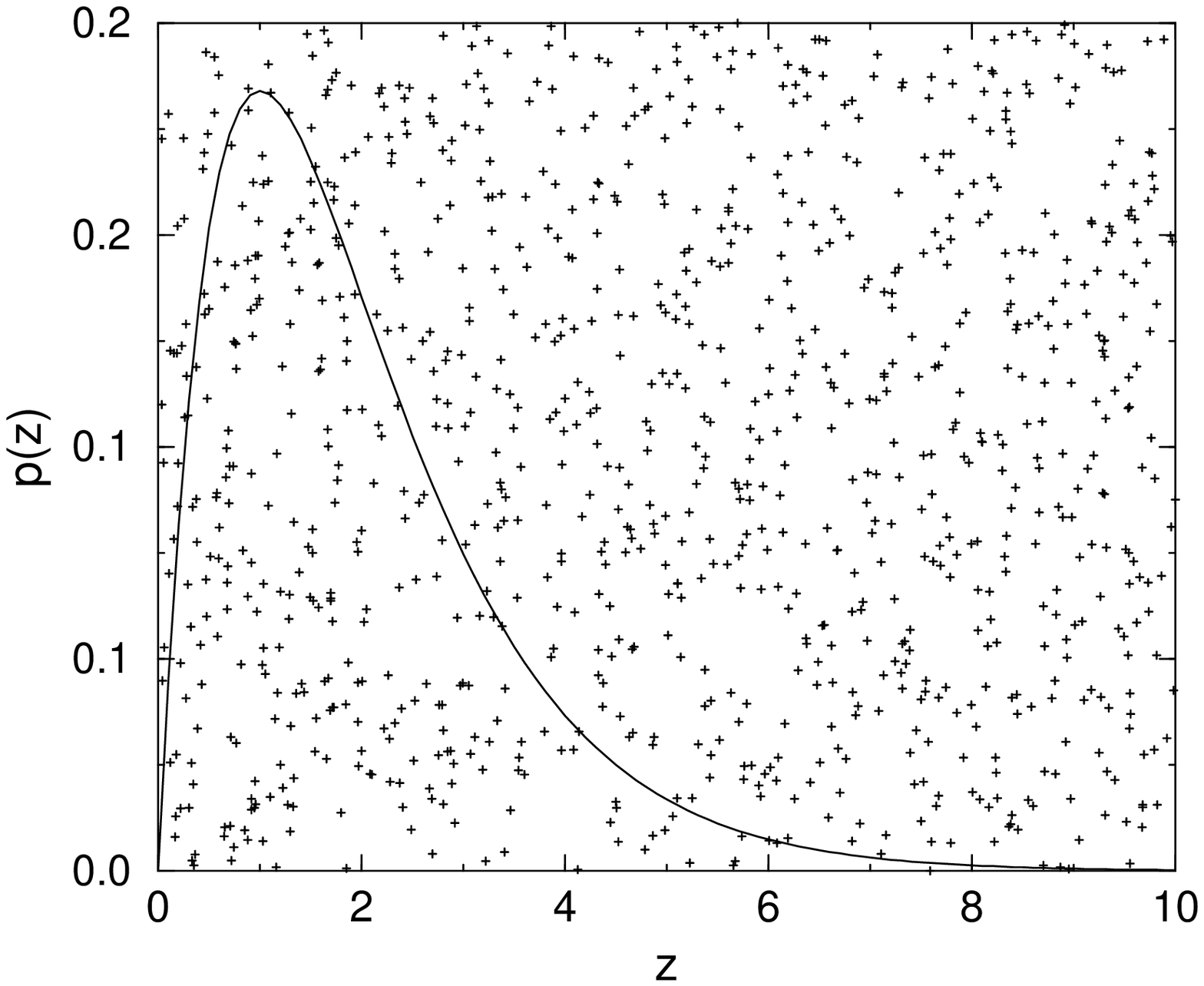}}
\caption{The rejection method: points ($x,y$) are scattered uniformly over a
  bounded rectangle. The probability that $y\le p(x)$ is proportional
  to $p(x)$.}
\label{fig:rejection}
\end{center}
\end{figure}

The {\em rejection method}\/, which is presented in this section,
 works for random variables where the probability
distribution $p(z)$ fits into a box $[x_0,x_1)\times [0,z_{\max})$,
i.e.\ $p(z)=0$ for $z\not\in [x_0,x_1]$ and $p(z)\le z_{\max}$. The
basic idea of generating a random number distributed according to $p(z)$ is
to generate random pairs ($x,y$), which are distributed uniformly in
$[x_0,x_1]\times [0,z_{\max}]$ and accept only those values $x$ where 
$y\le p(x)$ holds, i.e.\ the pairs which are located below $p(x)$, see
Fig.\ \ref{fig:rejection}. Therefore, the probability that $x$ is drawn is
proportional to $p(x)$, as desired. The algorithm for the rejection
method is:

\clearpage
\begin{algorithm}{rejection\_method($z_{\max}, p$)}
\>  $found:=$ {\bf false};\\
\> {\bf while} {\bf not} $found$ {\bf do}\\
\> {\bf begin}\\
\>\> $u_1:=$ random number in $[0,1)$;\\
\>\> $x:=x_0+(x_1-x_0)\times u_1$;\\
\>\> $u_2:=$ random number in $[0,1)$;\\
\>\> $y:=z_{\max}\times u_2$;\\
\>\> {\bf if} $y\le p(x)$ {\bf then}\\
\>\>\> $found:=$ {\bf true};\\
\> {\bf end};\\
\> {\bf return}(x);\\ 
\end{algorithm}

The rejection method always works if the probability density is boxed,
but it has the drawback that more random numbers have to be generated
than can be used.

In case neither the distribution function can be inverted nor the
probability fits into a box, special methods have to be applied. As an
example  a method for generating random numbers distributed
according to a Gaussian distribution is considered. Other methods and
examples of how different techniques can be combined, are collected in
Ref. \cite{PRA-morgan1984}.
\index{rejection method|)}

\subsection{The Gaussian Distribution}
\index{Gaussian distribution|(ii}
\index{distribution!Gaussian|(ii}

\begin{figure}[ht]
\begin{center}
\scalebox{0.5}{\includegraphics{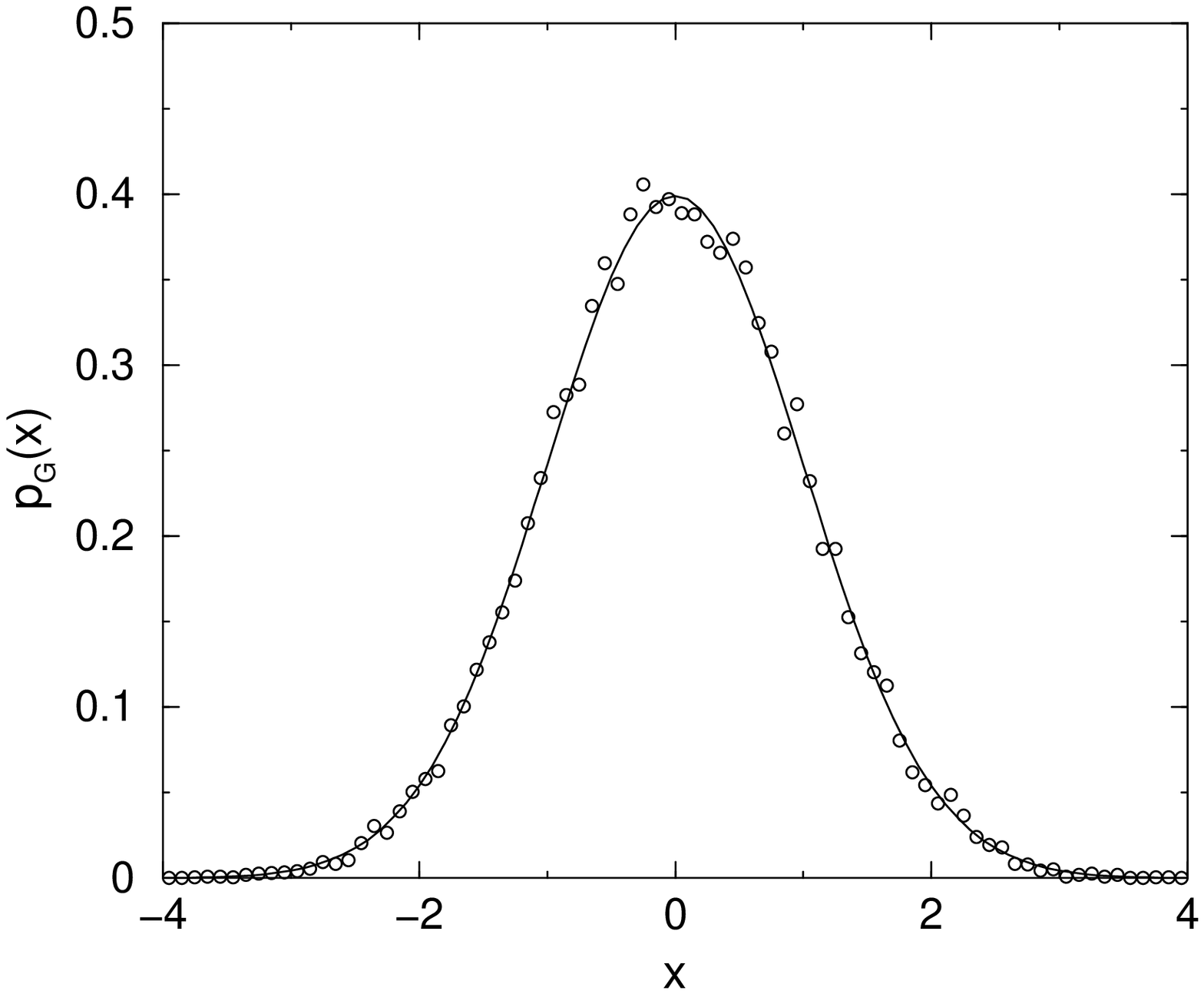}}
\caption{Gaussian distribution with zero mean and unit width. The
  circles represent a histogram obtained from $10^4$ values drawn with
  the Box-M\"uller method.}
\label{fig:Gauss}
\end{center}
\end{figure}

The probability density \index{probability density}
for the Gaussian distribution with mean $m$ and width
$\sigma$ is (see also Fig.\ \ref{fig:Gauss})
\begin{equation}
  p_G(z)= \frac{1}{\sqrt{2\pi}\sigma}\exp\left(\frac{(z-m)^2}{2\sigma^2}\right)
\label{eq:Gauss}
\end{equation}
It is, apart from uniform distributions, the most
common distribution being applied in simulations.

Here, the case of a normal distribution \index{normal distribution}
\index{distribution!normal}
($m=0,\;\sigma=1$) is considered. If you want to realize the general case, you
have to draw a normally distributed number $z$ and then use $\sigma
z+m$ which is distributed as desired.

Since the normal distribution extends over an infinite interval and
cannot be inverted, the methods from above are not applicable.
The simplest technique to generate random numbers distributed
according to a normal distribution makes use of the central limit
theorem. \index{central limit theorem} \index{theorem!central limit}
It tells us that any sum of $N$ independently distributed
random variables $u_i$ (with mean $m$ and variance $v$) will converge
to a Gaussian distribution with mean $Nm$ and variance $Nv$. If 
again $u_i$ is taken take to be uniformly distributed in $[0,1)$ (which has
mean $m=0.5$ and variance $v=1/12$), one can choose $N=12$ and
$Z=\sum_{i=1}^{12} u_i-6$ will be distributed approximately
normally. The drawback of this method is that 12 random numbers are
needed to generate one final random number and that values
larger than 6 never appear. 

In contrast to this technique the {\em Box-M\"uller method}\/ is
exact. You need two uniformly in $[0,1)$ distributed random variables
$U_1,U_2$ to generate two independent normal variables
$N_1,N_2$.  This can be achieved by setting
\begin{eqnarray*}
  N_1 & = & \sqrt{-2 \log (1-u_1)} \cos(2\pi u_2) \\
  N_2 & = & \sqrt{-2 \log (1-u_1)} \sin(2\pi u_2) 
\end{eqnarray*}
A proof that $N_1$ and $N_2$ are indeed distributed according to
(\ref{eq:Gauss}) can be found in Refs. \cite{PRA-numrec1995,PRA-morgan1984},
where also other methods for generating Gaussian random numbers, some even
more efficient, are
explained. A method which is based on the simulation of particles in a
box is explained in Ref. \cite{PRA-fernandez1999}. 
In Fig.\ \ref{fig:Gauss} a histogram of $10^4$ random
numbers drawn with the Box-M\"uller method is shown.
\index{random number generator|)}
\index{Gaussian distribution|)}
\index{distribution!Gaussian|)}

\section{Tools for Testing}
\label{sec-testing}
\index{debugging!tools|(}
\index{testing!tools|(}

In Sec.\ \ref{sec-engineering} the
importance of thorough testing has already been stressed. 
Here three useful tools are
presented  which significantly assist in facilitating the debugging 
process. Please note
again that the tools run under UNIX/Linux operating systems. Similar
programs are available for other operating systems as well.
The tools covered here are {\em gdb}\/, a source-code debugger, {\em
  ddd}\/, a graphic front-end to gdb, and {\em checkergcc}\/, which finds bugs
resulting from bad memory management.

\subsection{\em 
gdb}
\label{sec-gdb}
\index{gdb@{\em{}gdb}|(}

The {\em gdb} {gnu debugger} tool is a {\em source code debugger}\/. 
\index{source-code debugger} \index{debugging} Its main
purpose is that you can watch the execution of your code. You can stop
the program at arbitrarily chosen points by setting {\em breakpoints}\/
at lines or subroutines in the source code,
inspect variables/data structures, change them and let the program
continue (e.g.\ line by line). 
Here some examples for the most basic operations are given, 
detailed instructions can be obtained within the program via the {\tt
  help} command.

As an example of how to debug, please consider the following little
program {\tt gdbtest.c}:

\begin{verbatim}
#include <stdio.h>
#include <stdlib.h>

int main(int argc, char *argv[])
{
    int t, *array, sum = 0;

    array = (int *) malloc (100*sizeof(int));
    for(t=0; t<100; t++)
        array[t] = t;
    for(t=0; t<100; t++)
        sum += array[t];
    printf("sum= %d\n", sum);
    free(array);
    return(0);
}
\end{verbatim}

When compiling the code you have to include the option {\tt-g} to
allow debugging:
\begin{verbatim}
cc -o gdbtest -g gdbtest.c
\end{verbatim}
The debugger is invoked using {\tt gdb} {\it $<$programname$>$}, i.e.
\begin{verbatim}
gdb gdbtest
\end{verbatim}

Now you can enter commands, e.g.\ list the source code of the program
via the {\tt list} command, it is sufficient to enter just {\tt{}l}.
 By default always ten lines at the current
position are printed. Therefore, at the beginning the first ten lines
are shown (the first line shows the input, the other lines state the
answer of the debugger)

\clearpage
\begin{verbatim}
(gdb) l
1       #include <stdio.h>
2       #include <stdlib.h>
3
4       int main(int argc, char *argv[])
5       {
6           int t, *array, sum = 0;
7
8           array = (int *) malloc (100*sizeof(int));
9           for(t=0; t<100; t++)
10              array[t] = t;
\end{verbatim}
When entering the command again the next ten lines are
listed. Furthermore, you can refer to program lines of the code in the form
{\tt list} {\it $<$from$>$, $<$to$>$} or to subroutines by typing {\tt list}
{\it $<$name of subroutine$>$}. More information can be obtained by typing
{\tt help list}.

To let the execution stop at a specific line one can use the {\tt
  break} command (abbreviation {\tt{}b}). To stop the program {\em
  before}\/ line 11 is executed, one enters
\begin{verbatim}
(gdb) b 11
Breakpoint 1 at 0x80484b0: file gdbtest.c, line 11.
\end{verbatim}
Breakpoints can be removed via the {\tt delete} command. All current
breakpoints are displayed by entering {\tt info break}.

To start the execution of the program, one enters {\tt run} or
just {\tt{}r}. As requested before, the program will stop at line 11:
\begin{verbatim}
(gdb) r
Starting program: gdbtest 

Breakpoint 1, main (argc=1, argv=0xbffff384) at gdbtest.c:11
11          for(t=0; t<100; t++)
\end{verbatim}
Now you can inspect for example the content of variables via the
{\tt{}print} command:
\begin{verbatim}
(gdb) p array
$1 = (int *) 0x8049680
(gdb) p array[99] 
$2 = 99
\end{verbatim}
To display the content of a variable permanently, the {\tt display}
command is available.
You can change the content of variables via the {\tt set} command
\begin{verbatim}
(gdb) set array[99]=98
\end{verbatim}
You can continue the program at each stage by typing
{\tt next}, then just the next source-code line is executed:
\begin{verbatim}
(gdb) n
12              sum += array[t];
\end{verbatim}
Subroutines are regarded as one source-code line as well. If you want
to debug the subroutine in a step-wise  manner 
as well you have to enter the {\tt
  step} command.
By entering {\tt continue}, the execution is continued until the next
breakpoint, a severe error, 
or the end of the program is reached, please note the the
output of the program appears in the {\em gdb} window as well:
\begin{verbatim}
(gdb) c
Continuing.
sum= 4949

Program exited normally.
\end{verbatim}
As you can see, the final value (4949) the program prints is affected
by the change of the variable {\tt array[99]}.

The above given commands are sufficient for most of the
standard debugging tasks. For more specialized cases {\em gdb} offers many
other commands, please have a look at the documentation \cite{PRA-gnu}.
\index{gdb@{\em{}gdb}|)}

\subsection{\em ddd}

\begin{figure}[ht]
\begin{center}
\scalebox{0.6}{\includegraphics{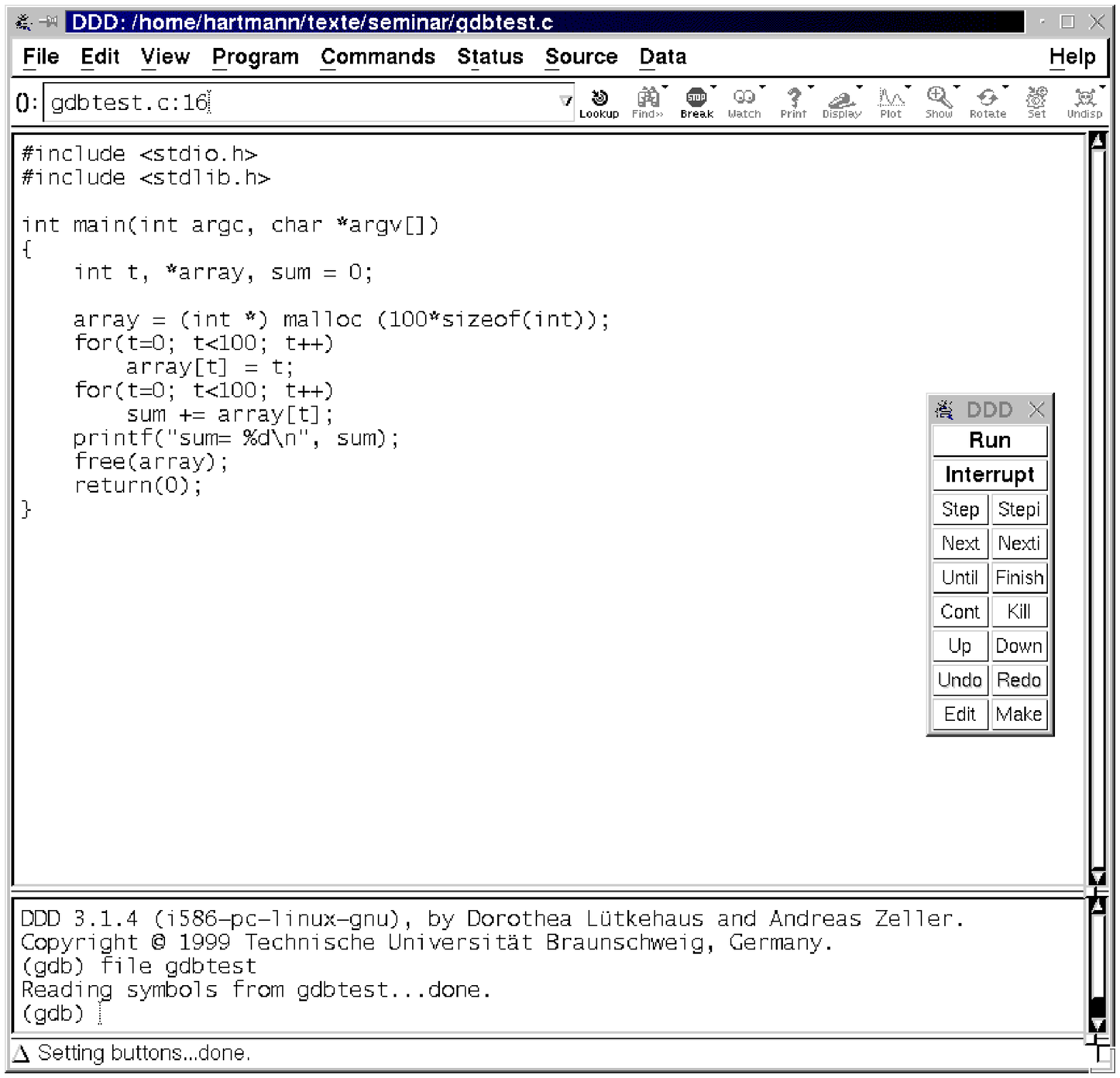}}
\caption{The data display debugger (ddd). In the main window the
  source code is shown. Commands can be invoked via a mouse or by
  entering them into the lower part of the window.}
\label{fig:ddd}
\end{center}
\end{figure}

Some users may find graphical user interfaces more convenient. For this
reason there exists a graphical front-end to the gdb, the {\em data
  display debugger (ddd)\/}. \index{data display debugger ({\em{}ddd})|(}
On UNIX operating systems it is just invoked
by typing ddd (see also {\em man}\/ page for options). Then a nice windows pops
up, see Fig.\ \ref{fig:ddd}. The lower part of the window is an
ordinary {\em gdb}\/ interface, several other windows are available. 
By typing {\tt file} {\it $<$program$>$} you can
load a program into the debugger. Then the source code is shown in the
main window of the debugger. All {\em gdb}\/ commands are available,
the most important ones can be entered via menus or buttons using the
mouse. For example to set a breakpoint it is sufficient to place the
cursor in a source-code line in the main ddd window and click on the
{\it break} button. A good feature is that the content of a variable is
shown when moving the mouse onto it.
For more details, please consult the online help of ddd.

\index{data display debugger ({\em{}ddd})|)}

\subsection{\em checkergcc}
\index{checkergcc|(}

Most program bugs are revealed by systematically running the program and
cross-checking with the expected results. But other errors seem to
appear in a rather irregular and unpredictable fashion. 
Sometimes a program runs without a
problem, in other cases it crashes with a {\tt Segmentation fault} 
\index{segmentation fault} at
rather puzzling locations in the code. Very often a bad
memory management is the cause of such a behavior. Writing beyond the
boundaries of an array, \index{array} reading uninitialized memory locations or
addressing data which has been freed already are the most common
bugs of this class. Since the operating system organizes the memory in a
different way each time a program is run, it is rather unpredictable
whether these errors become apparent or not. Furthermore it is very
hard to track them down, because the effect of such errors most of the 
time becomes visible at positions different from where the error has 
occurred. 

As an example, the case where it is written beyond the 
boundary of an array is considered. 
If in the heap, where all dynamically allocated
memory is taken from, at the location behind the array another variable
is stored, it will be overwritten in this case. Hence, the error 
becomes visible the
next time the other variable is read. On the other hand, if the memory
block behind the array is not used, the program may run that time 
without any problems. Unfortunately, the programmer is not able to
influence the memory management directly.

To detect such types of nasty bugs, one can take advantage of several
tools. A list of free and commercial tools can
be found in Ref. \cite{PRA-memorytools}. Here  {\em
  checkergcc}\/ is considered, which  is
 a very convenient tool and freely available. It works under
 UNIX \index{UNIX} and is included by compiling everything with {\tt
   checkergcc} instead of {\tt cc} or {\tt gcc}. Unfortunately, the
 current version does not have full support for C++, but you should try it
 on your own project. The checkergcc compiler
 replaces all memory allocations/deallocations \index{memory allocation}
and accesses by its
 own routines. Any access to non-authorized memory locations is
 reported, regardless of the positions of other variables in the memory
 area (heap).

As an example, the program from Sec.\ \ref{sec-gdb} is considered, which
is slightly modified; the memory block allocated for the array is now slightly
too short (length 99 instead of 100):
\begin{verbatim}
#include <stdio.h>
#include <stdlib.h>

int main(int argc, char *argv[])
{
    int t, *array, sum = 0;

    array = (int *) malloc (99*sizeof(int));
    for(t=0; t<100; t++)
        array[t] = t;
    for(t=0; t<100; t++)
        sum += array[t];
    printf("sum= %d\n", sum);
    free(array);
    return(0);
}
\end{verbatim}

The program is compiled via
\begin{verbatim}
checkergcc -o gdbtest -g gdbtest.c
\end{verbatim}
Starting the program produces the following output, the program
terminates normally:
\begin{verbatim}
Sisko:seminar>gdbtest
Checker 0.9.9.1 (i686-pc-linux-gnu) Copyright (C) 1998 Tristan Gingold.
This program has been compiled with 'checkergcc' or 'checkerg++'.
Checker is a memory access detector.
Checker is distributed in the hope that it will be useful,
but WITHOUT ANY WARRANTY; without even the implied warranty of
MERCHANTABILITY or FITNESS FOR A PARTICULAR PURPOSE.  See the GNU
General Public License for more details.
For more information, set CHECKEROPTS to '--help'
From Checker (pid:30448): `gdbtest' is running 

From Checker (pid:30448): (bvh) block bounds violation in the heap.
When Writing 4 byte(s) at address 0x0805fadc, inside the heap (sbrk).
0 byte(s) after a block (start: 0x805f950, length: 396, mdesc: 0x0).
The block was allocated from:
        pc=0x080554f9 in chkr_malloc at stubs-malloc.c:57
        pc=0x08048863 in main at gdbtest.c:8
        pc=0x080555a7 in this_main at stubs-main.c:13
        pc=0x40031c7e in __divdi3 at stubs/end-stubs.c:7
        pc=0x08048668 in *unknown* at *unknown*:0
Stack frames are:
        pc=0x080489c3 in main at gdbtest.c:10
        pc=0x080555a7 in this_main at stubs-main.c:13
        pc=0x40031c7e in __divdi3 at stubs/end-stubs.c:7
        pc=0x08048668 in *unknown* at *unknown*:0
From Checker (pid:30448): (bvh) block bounds violation in the heap.
When Reading 4 byte(s) at address 0x0805fadc, inside the heap (sbrk).
0 byte(s) after a block (start: 0x805f950, length: 396, mdesc: 0x0).
The block was allocated from:
        pc=0x00000063 in *unknown* at *unknown*:0
        pc=0x08048863 in main at gdbtest.c:8
        pc=0x080555a7 in this_main at stubs-main.c:13
        pc=0x40031c7e in __divdi3 at stubs/end-stubs.c:7
        pc=0x08048668 in *unknown* at *unknown*:0
Stack frames are:
        pc=0x08048c55 in main at gdbtest.c:12
        pc=0x080555a7 in this_main at stubs-main.c:13
        pc=0x40031c7e in __divdi3 at stubs/end-stubs.c:7
        pc=0x08048668 in *unknown* at *unknown*:0
\end{verbatim}
Two errors are reported, each message starts with ``{\tt{}From
  checker}''. Both errors consist of accesses to an array beyond the border
({\tt block bound violation}). For each error both the location in the
source code where the memory has been allocated and the location 
where the error
occurred ({\tt Stack frames}) \index{stack!frame}
are given. In both cases the error is
concerned with what was allocated at line 8 ({\tt pc=0x08048863 in main
  at gdbtest.c:8}). The bug appeared during the loops over the array,
when the array is initialized (line 10) and read out (line 12).

Other common types of errors are memory leaks. \index{memory leak}
They appear when a
previously used block of memory has been forgotten to be freed
again. Assume that this happens in a subroutine which is called frequently
in a program. You can imagine that you will quickly run out of
memory. Memory leaks are not detected using checkergcc by default. 
This kind of test can be turned on by
setting a special environment variable {\tt CHECKEROPTS},
which controls the behavior of {\tt checkergcc}. To enable checking for
memory leaks at the end of the execution, one has to set
\begin{verbatim}
export CHECKEROPTS="-D=end"
\end{verbatim}
Let us assume that the bug from above is removed and instead the {\tt
  free(array);} command at the end of the program is omitted. After
compiling with {\tt checkergcc}, running the program results in:
\begin{verbatim}
From Checker (pid:30900): `gdbtest' is running 

sum= 4950
Initialization of detector...
Searching in data
Searching in stack
Searching in registers
From Checker (pid:30900): (gar) garbage detector results.
There is 1 leak and 0 potential leak(s).
Leaks consume 400 bytes (0 KB) / 132451 KB.
( 0.00% of memory is leaked.)
Found 1 block(s) of size 400.
Block at ptr=0x805f8f0
        pc=0x08055499 in chkr_malloc at stubs-malloc.c:57
        pc=0x08048863 in main at gdbtest.c:8
        pc=0x08055547 in this_main at stubs-main.c:13
        pc=0x40031c7e in __divdi3 at stubs/end-stubs.c:7
        pc=0x08048668 in *unknown* at *unknown*:0
\end{verbatim}
Obviously, the memory leak has been found. Further information on the
various features of checkergcc can be found in
Ref. \cite{PRA-checkergcc}. A last hint: you should always test a program
with a memory checker, even if everything seems to be fine.
\index{debugging!tools|)}
\index{testing!tools|)}
\index{checkergcc|)}

\section{Evaluating Data}
\index{data!analysis|(}

To analyze and plot data, several commercial and non-commercial programs are
available. Here three free programs are discussed, 
{\em gnuplot}\/, {\em xmgr}\/ and
{\em fsscale}\/.  {\em Gnuplot}\/
is small, fast, allows two- and three-dimensional curves to be generated and
to fit arbitrary functions to the data. 
On the other hand {\em xmgr}\/ is more flexible and produces better output. 
It is recommended that
{\em gnuplot}\/ is used  for viewing and fitting data online, 
while {\em xmgr}\/
is to be preferred for producing figures to be shown in talks or publications.
The program fsscale has a special purpose. It is very convenient for performing
finite-size scaling plots. 

First,  {\em gnuplot}\/ and {\em xmgr}\/ are introduced with respect to drawing
figures. In the next subsection, data fitting is covered. Finally,
it is shown how finite-size scaling plots can be created. In all three
cases only very small examples can be presented. They should serve
just as a motivation to study the documentation, then you will learn about the
manifold potential the programs offer.

\subsection{Data Plotting} 
\index{gnuplot@{\em{}gnuplot}|(ii}
The program {\em gnuplot}\/ is invoked by entering {\tt gnuplot} in a
shell, for a complete manual see Ref. \cite{PRA-texinfo}. 
As always, our examples refer to a UNIX window system like X11, 
but the program is available for
almost all operating systems. After startup, the prompt (e.g.\ {\tt
  gnuplot$>$}) appears
and the user can enter commands in textual form, results are shown in
windows or are written into files. 
Before giving an example, it should be
pointed out that gnuplot {\em scripts}\/ \index{gnuplot@{\em{}gnuplot}!script}
 can be generated by simply writing the commands
into a file, e.g.\ {\tt command.gp}, and calling {\tt gnuplot
  command.gp}.

\begin{figure}[ht]
\begin{center}
\scalebox{0.6}{\includegraphics{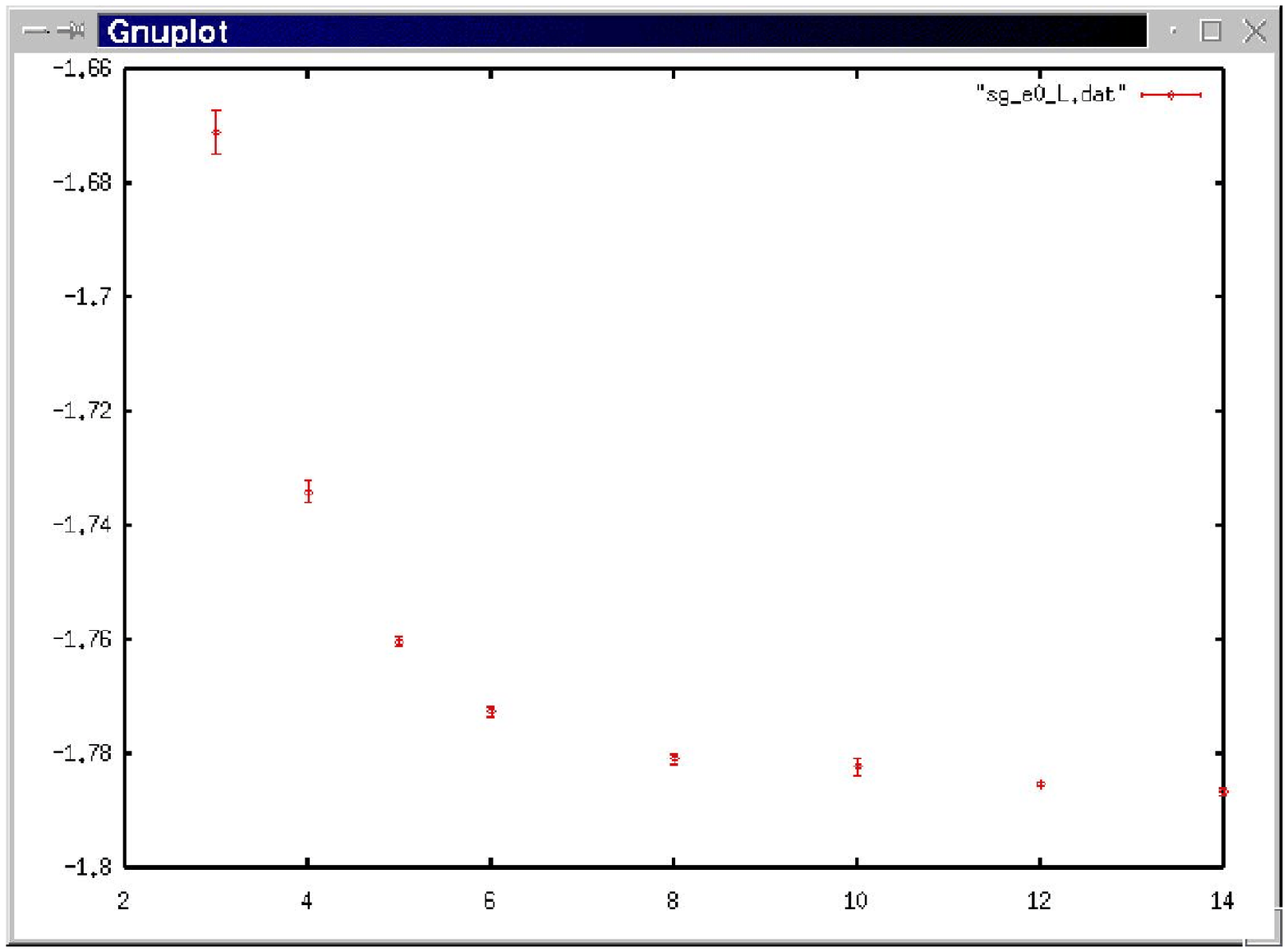}}
\caption{Gnuplot window showing the result of a plot command.}
\label{fig:gnuplotA}
\end{center}
\end{figure}

The typical case is that you have a data file of $x-y$ data and you want
to plot the figure. Your file might look like this, it is the
ground-state energy \index{ground state!energy}
of a three-dimensional $\pm J$ spin glass \index{spin glass}
as a function of the linear
system size $L$. The filename is {\tt sg\_e0\_L.dat}. The first column
contains the $L$ values, the second the energy values and the third the
standard error of the energy, please note that lines starting with
``{\tt{}\#}'' are comment lines which are ignored on reading:
\begin{verbatim}
# ground state energy of +-J spin glasses
# L    e_0   error
  3 -1.6710 0.0037
  4 -1.7341 0.0019
  5 -1.7603 0.0008
  6 -1.7726 0.0009
  8 -1.7809 0.0008
 10 -1.7823 0.0015
 12 -1.7852 0.0004
 14 -1.7866 0.0007
\end{verbatim}
To plot the data enter
\begin{verbatim}
gnuplot> plot "sg_e0_L.dat" with yerrorbars
\end{verbatim}
which can be abbreviated as {\tt p "sg\_e0\_L.dat" w e}. Please do not
forget the quotation marks around the file name. Next, a window
pops up, showing the result, see Fig.\ \ref{fig:gnuplotA}.

For the {\tt plot} command many options and styles are available, e.g.\ {\tt
  with lines} produces lines instead of symbols. It is possible to
read files with multi columns via the {\tt using} option, e.g.
\begin{verbatim}
gnuplot> plot "test.dat" using 1:4:5 w e
\end{verbatim}
displays the fourth column as a function of the first, with error bars
given by the 5th column. Among other options, it is possible to
redirect the output, for example to an encapsulated
 postscript file \index{postscript@{\em{}postscript}}
(by setting {\tt
  set terminal postscript} and redirecting the output {\tt set output
   "test.eps"}).  Also several files can be combined into
one figure. You can set axis labels of the figure by
typing e.g.\ {\tt set xlabel "L"}, which becomes active when the next
plot command is executed.  Online help on the plot command and its
manifold options is available  
via entering {\tt help plot}. Also three-dimensional plotting is possible
using the {\tt splot} command (enter {\tt help splot} to obtain more
information).  
For a general introduction you can type just {\tt help}.
Since {\em gnuplot}\/ commands can be entered very quickly, you should use it
for online viewing data and fitting (see Sec.\ \ref{sec:fitting}).
\index{gnuplot@{\em{}gnuplot}|)}

\begin{figure}[ht]
\begin{center}
\scalebox{0.6}{\includegraphics{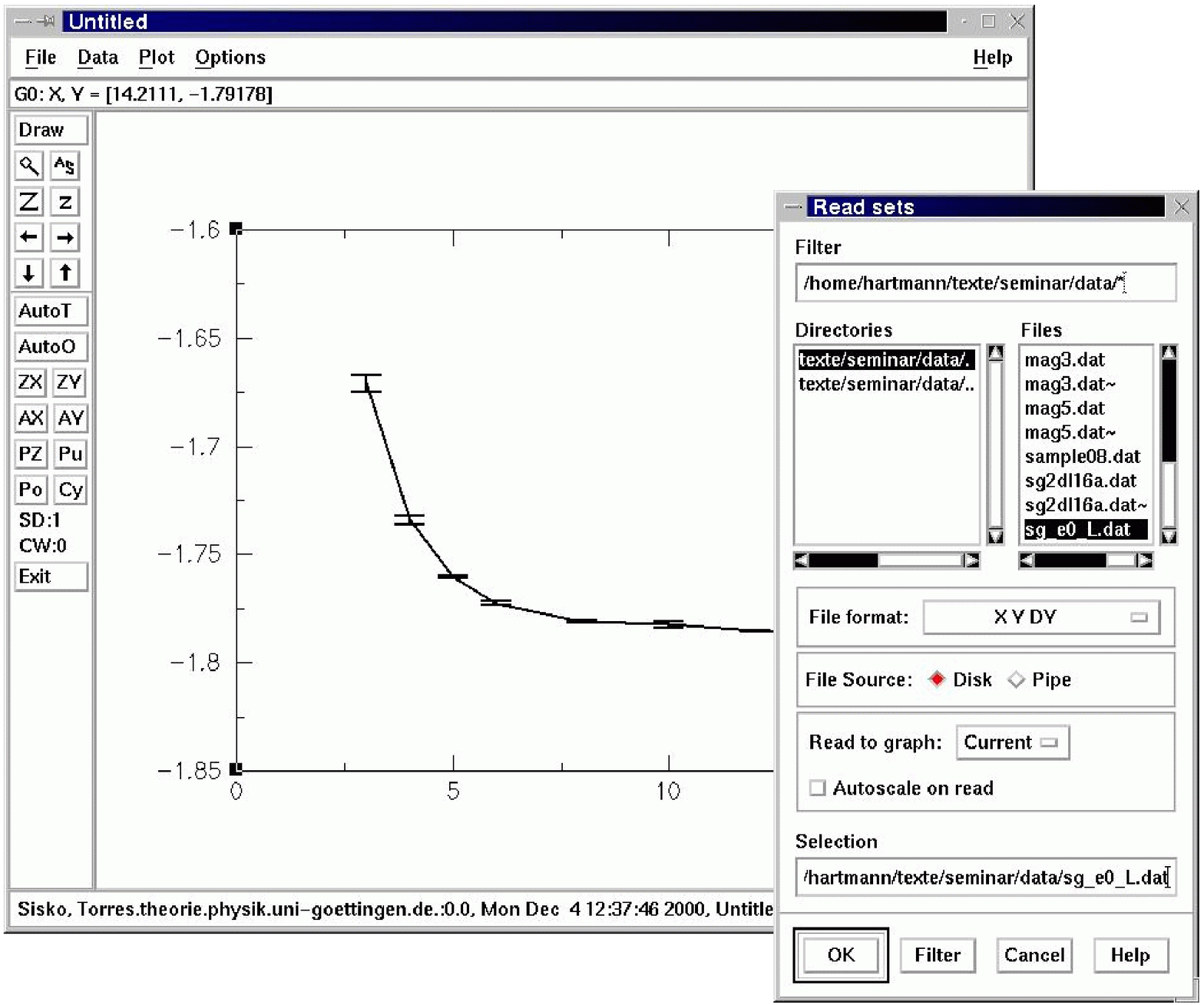}}
\caption{The {\em xmgr}\/ program, just after a data file has been loaded, and
  the AS button has been pressed to adjust the figure range automatically.}
\label{fig:xmgr:sample}
\end{center}
\end{figure}

\index{xmgr@{\em{}xmgr}|(}
The {\em xmgr}\/ (x motiv graphic) program is much more powerful than
{\em gnuplot}\/ and produces nicer output, commands are issued by clicking on
menus and buttons. On the other hand its handling is a little bit
slower and the program has the tendency to fill your screen with
windows. To create a similar plot to that above, you have to go (after
staring it by typing {\tt xmgr} into a shell) to the files menu and
choose the read submenu and the sets subsubmenu. Then a file selection
window will pop up and you can choose the data file to be loaded. The
situation is shown in Fig.\ \ref{fig:xmgr:sample}.

The {\em xmgr}\/ program offers almost every feature you can imagine for
two-di\-men\-sio\-nal data plots, including multiple plots, fits, many
styles for lines, symbols, bar charts etc. Also you can create
manifold types of labels or legends and it is possible to add elements
like strings, lines or other geometrical objects in the plot. For more
information, please consult the online help. 
\index{xmgr@{\em{}xmgr}|)}

\subsection{Curve Fitting}
\label{sec:fitting} 
\index{curve fitting|(} \index{fitting|(}

Both programs presented above, {\em gnuplot}\/ and {\em xmgr}\/, offer fitting of
arbitrary functions. It is advisable to use {\em gnuplot}\/, since it offers 
a higher flexibility for that purpose and gives you more information
useful to estimate the quality of a fit. 

As an example, let us suppose that you want to fit an algebraic function
of the form $f(L)=e_\infty +aL^b$ to the data set  of the file {\tt
  sg\_e0\_L.dat} shown above. First, you have to define the function and
supply some roughly (non-zero) estimations for the unknown parameters,
please note that the exponential operator is denoted by {\tt **} and the
standard argument for a function definition is {\tt x}, but this
depends only on your choice:
\begin{verbatim}
gnuplot> f(x)=e+a*x**b
gnuplot> e=-1.8
gnuplot> a=1
gnuplot> b=-1
\end{verbatim}
The actual fit is performed via the {\tt fit} command. The program
uses the nonlinear least-squares
 Marquardt-Levenberg algorithm \cite{PRA-numrec1995}, which allows
 a fit according to almost all arbitrary functions. To issue the command,
 you have to state the fit function, the data set and the parameters
 which are to be adjusted. For our example you enter:

\begin{verbatim}
gnuplot> fit f(x) "sg_e0_L.dat" via e,a,b
\end{verbatim}

Then {\em gnuplot}\/ writes log information to the output describing the
fitting process. After the fit has converged it prints for the given example:

\begin{verbatim}
After 17 iterations the fit converged.
final sum of squares of residuals : 7.55104e-06
rel. change during last iteration : -2.42172e-09

degrees of freedom (ndf) : 5
rms of residuals      (stdfit) = sqrt(WSSR/ndf)      : 0.00122891
variance of residuals (reduced chisquare) = WSSR/ndf : 1.51021e-06

Final set of parameters            Asymptotic Standard Error
=======================            ==========================

e               = -1.78786         +/- 0.0008548    (0.04781%)
a               = 2.5425           +/- 0.2282       (8.976%)
b               = -2.80103         +/- 0.08265      (2.951%)


correlation matrix of the fit parameters:

               e      a      b      
e               1.000 
a               0.708  1.000 
b              -0.766 -0.991  1.000 
\end{verbatim}
The most interesting lines are those where the results for your
parameters along with the standard error \index{error bar}
are printed. Additionally, the quality
of the fit can be estimated by the information provide in the three
lines beginning with ``{\tt degree of freedom}''. \index{degree!of freedom}
The first of these lines
states the number of degrees of freedom, which is just the number of
data points minus the number of parameters in the fit. The deviation
of the fit function $f(x)$ from the data points $(x_i,y_i\pm \sigma_i)$ 
($i=1,\ldots,N$) is given by $\chi^2=\sum_{i=1}^N
\left[\frac{y_i-f(x_i)}{\sigma_i}\right]^2$, 
which is denoted by WSSR in the {\em gnuplot}\/
output. A measure of the quality of the fit \index{quality of fit}
\index{fitting!quality of} is the probability $Q$
that the value of $\chi^2$ is worse than in the current fit, given the
assumption that the datapoints $y_i$ are Gaussian distributed with mean 
$f(x_i)$ and variance one \cite{PRA-numrec1995}. The larger the value of
$Q$, the better is the quality of the fit.
To calculate $Q$ you can use the little program {\tt Q.c} 
\clearpage
\begin{verbatim}
#include <stdio.h>
#include "nr.h"
int main(int argc, char **argv)
{
  float ndf, chi2_per_df;
  sscanf(argv[1], "%f", &ndf);
  sscanf(argv[2], "%f", &chi2_per_df);
  printf("Q=%e\n", gammq(0.5*ndf, 0.5*ndf*chi2_per_df));
  return(0);
}
\end{verbatim}
which uses the {\tt gammaq} function from {\em Numerical Recipes}\/
\cite{PRA-numrec1995}. The program is called in the form {\tt Q <ndf>
  <WSSR/ndf>}, which can be taken from the {\em gnuplot}\/ output.

\begin{figure}[ht]
\begin{center}
\scalebox{0.6}{\includegraphics{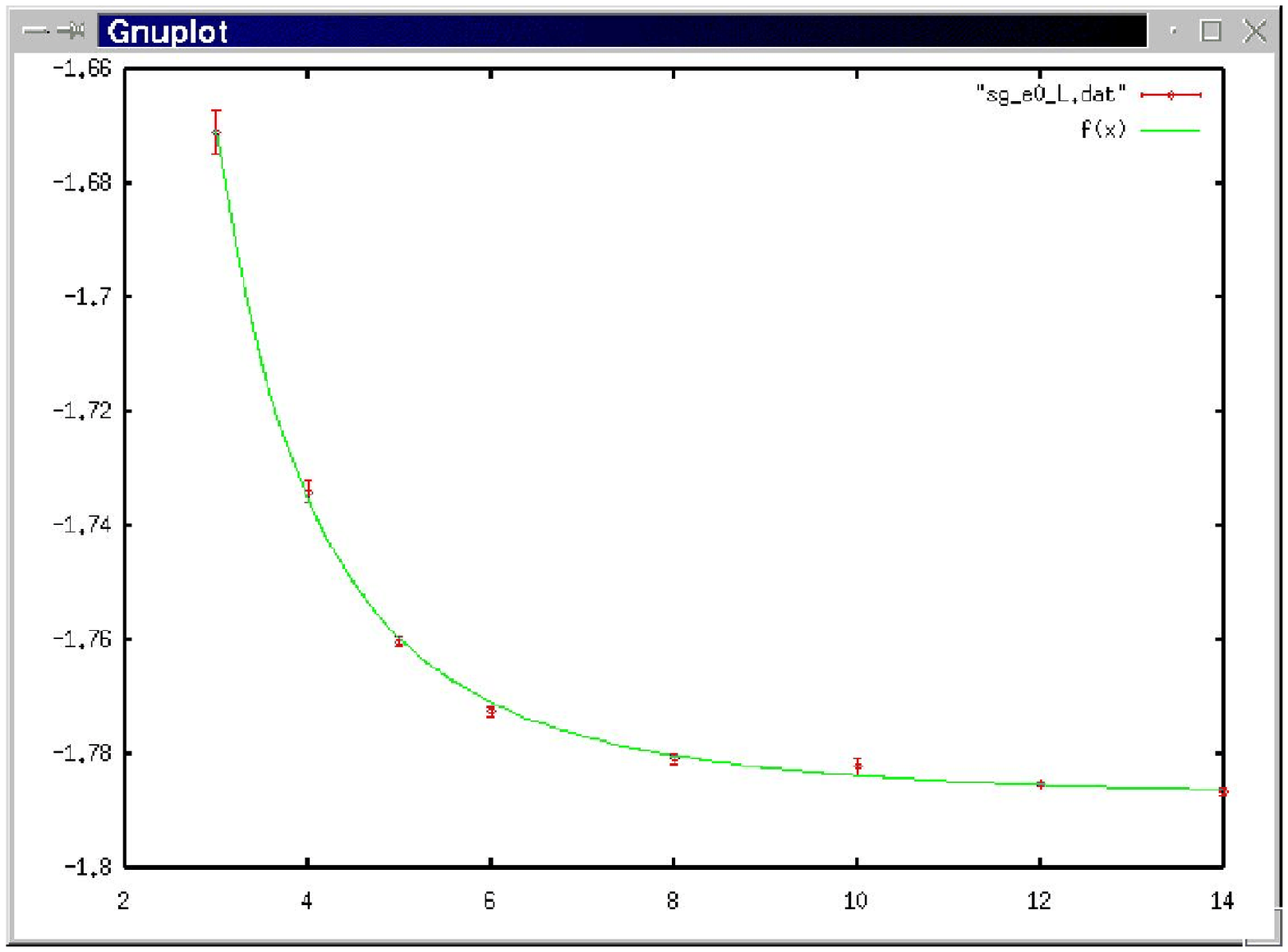}}
\caption{{\em Gnuplot}\/ window showing the result of a fit command along with
  the input data.}
\label{fig:gnuplotB}
\end{center}
\end{figure}

To watch the result of the fit along with the original data, just enter
\begin{verbatim}
gnuplot> plot "sg_e0_L.dat" w e, f(x)
\end{verbatim}
The result looks like that shown in Fig.\ \ref{fig:gnuplotB}

Please note that the
 convergence depends on the initial choice of the parameters. The
 algorithm may be trapped into a local minimum in case the parameters
 are too far away from the best values. Try the initial values
 {\tt e=1}, {\tt a=-3} and {\tt b=1}! 
Furthermore, not all function parameters have to be subjected to
the fitting. Alternatively, you can set some parameters to fixed
values and omit them from the list at the end of the {\em fit}\/
command. You should also know that
 in the example given above all data points enter into the
result with the same weight. You can tell the algorithm to consider
the error bars by typing {\tt fit f(x) "sg\_e0\_L.dat" using 1:2:3 via
  a,b,c}. Then, data points with larger error bars have less influence
on the results. More on how to use the {\tt fit} command can be
found out when entering {\tt help fit}.
\index{curve fitting|)} \index{fitting|)}

\subsection{Finite-size Scaling} 
\label{sec:practical:FSS}
\index{finite-size scaling|(}

Statistical physics describes the behavior of systems with many
particles. Usually, realistic system sizes cannot be simulated on
current computers. To circumvent this problem, the technique of {\em
  finite-size scaling}\/ has been invented, for an introduction see
e.g.\ Ref. \cite{PRA-cardy1996}. The basic idea is to simulate systems of
different sizes and extrapolate to the large volume limit. Here it is
shown how finite-size scaling can be performed with the help of {\em
  gnuplot}\/ \cite{PRA-texinfo} 
or with the special-purpose program {\em fsscale}\/ \cite{PRA-fsscale} 

\begin{figure}[ht]
\begin{center}
\scalebox{0.5}{\includegraphics{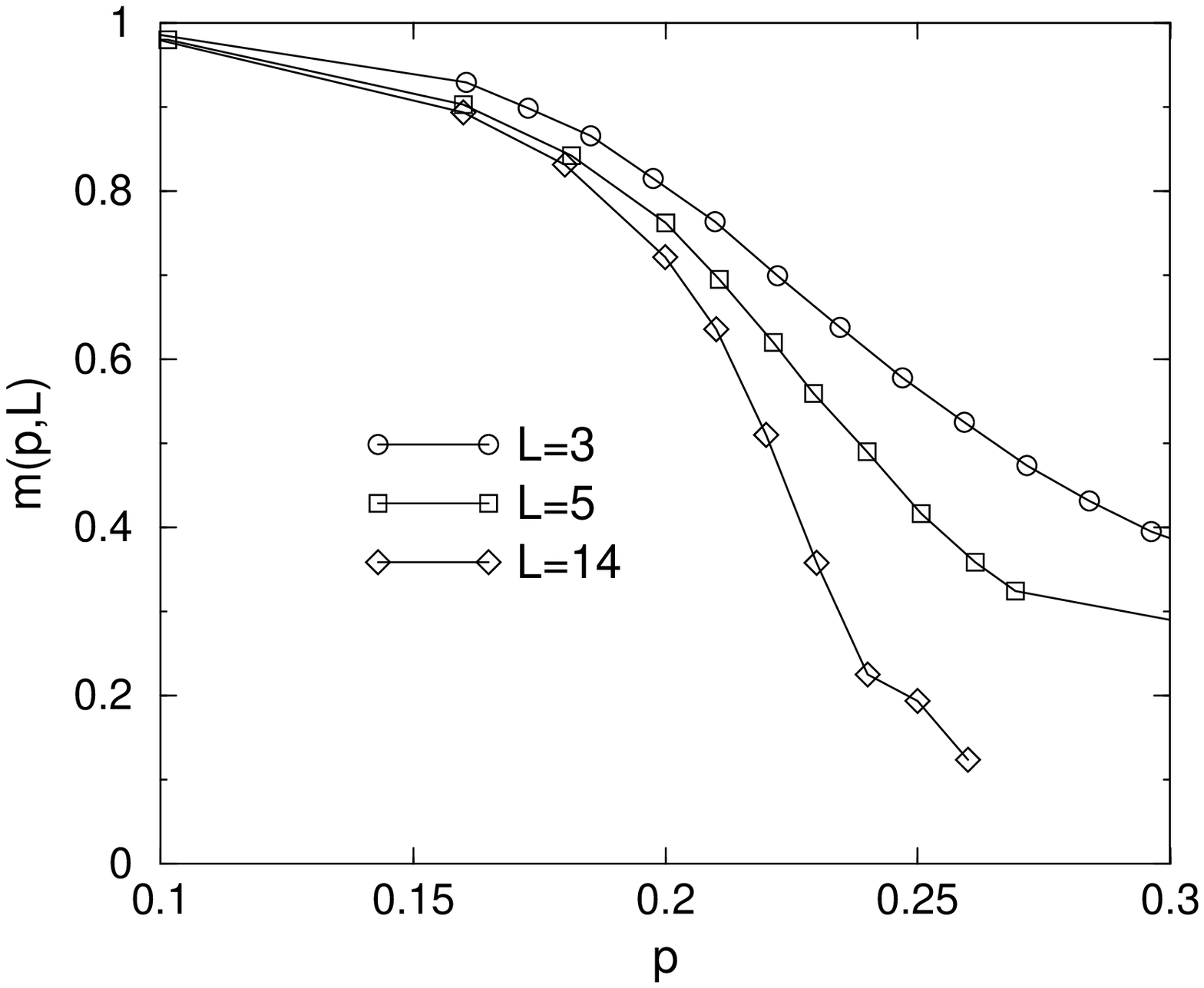}}
\caption{Average ground-state 
magnetization $m$ of a three-dimensional
$\pm J$ spin glass with  fractions $p$ of antiferromagnetic
bonds. Lines are guides to the eyes only.}
\label{fig:pmsg}
\end{center}
\end{figure}

As an example, the average ground-state 
magnetization $m$ \index{magnetization} of a three-dimensional
$\pm J$ spin glass \index{spin glass}
with  fractions $p$ of antiferromagnetic and $1-p$ of
ferromagnetic bonds is considered. 
For small values of $p$ the system is expected to
have a ferromagnetically ordered state. This can be observed 
in Fig.\ \ref{fig:pmsg}, where the results \cite{PRA-alex-threshold} for
different system sizes $L=3,5,14$ are shown.

\begin{figure}[ht]
\begin{center}
\scalebox{0.6}{\includegraphics{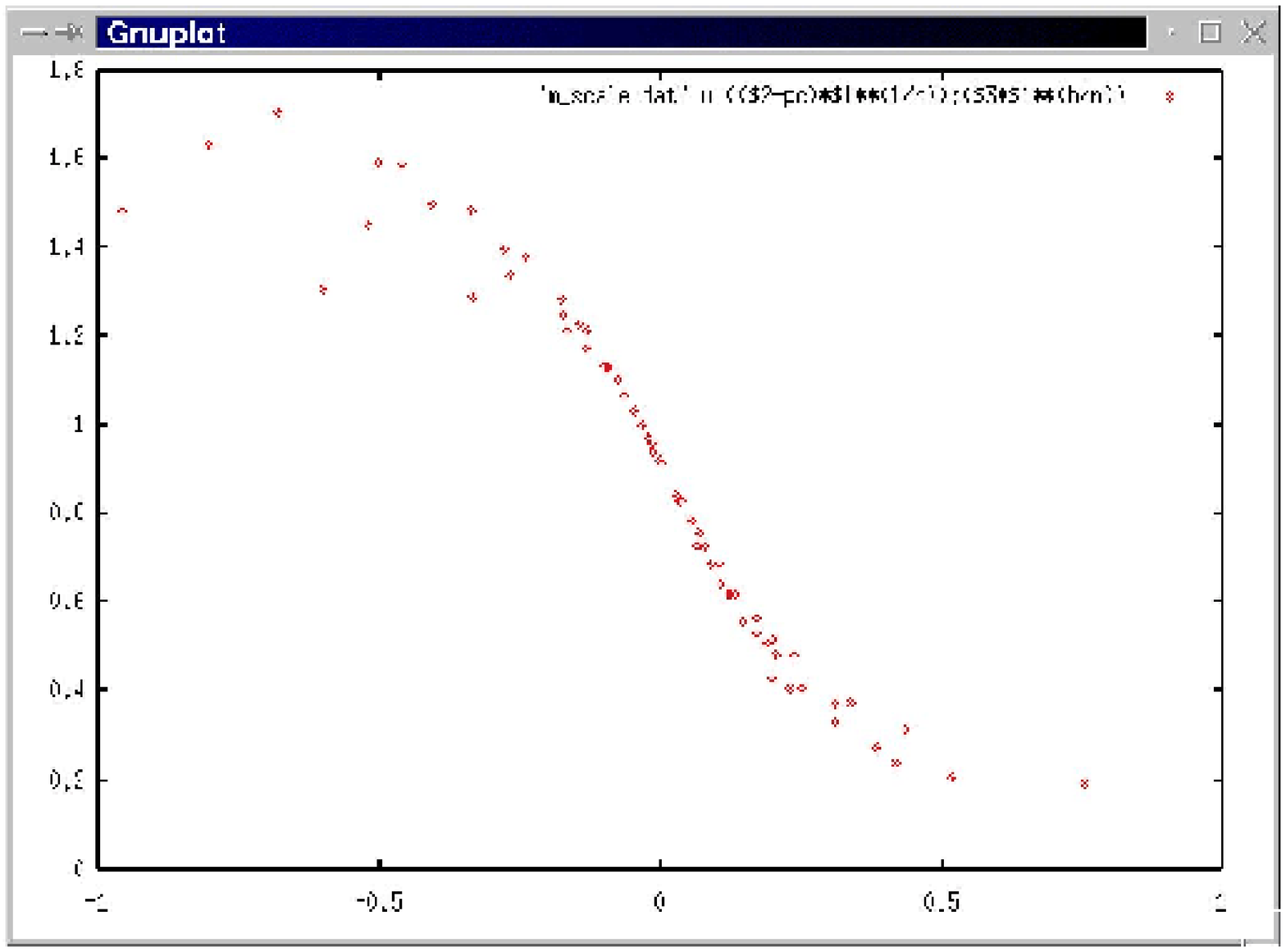}}
\caption{{\em Gnuplot}\/ output of a 
finite-size scaling plot. The ground-state magnetization of a
  three-dimensional $\pm J$ spin glass as a
  function of the concentration $p$ of the antiferromagnetic
  bonds is shown. For the fit, the parameters $p_{\rm c}=0.222,\beta=0.27$ and
  $\nu=1.1$ have been used.}
\label{fig:fssGnuplot}
\end{center}
\end{figure}

The critical concentration $p_{\rm c}$, where
the magnetization $m$ vanishes, and the critical behavior of $m$
near the transition are to be obtained.
From the theory of finite-size scaling, it is known that 
the average magnetization $m\equiv\langle M \rangle$ obeys 
the finite-size  scaling form \cite{PRA-binder-heermann}
\begin{equation}
m(p,L)=L^{-\beta/\nu}\tilde{m}(L^{1/\nu}(p-p_{\rm c})) \label{eq:fsscaleMag}
\end{equation}
where $\tilde{m}$ is a universal, i.e.\ non size-dependent, function.
The exponent $\beta$ characterizes the algebraic behavior of the
magnetization near $p_{\rm c}$, while the exponent $\nu$ describes the
divergence of the correlation length when $p_{\rm c}$ is approached. From
Eq.\ (\ref{eq:fsscaleMag}) you can see that when plotting
$L^{\beta/\nu}m(p,L)$ against $L^{1/\nu}(p-p_{\rm c})$ with correct
parameters $\beta,\nu$ the data points for different system sizes
should collapse onto a single curve. A good collapse can be obtained by
using the
values $p_{\rm c}=0.222$, $\nu=1.1$ and $\beta=0.27$. 
The determination of $p_{\rm c}$ and the exponents can be
performed via {\em
  gnuplot}\/. \index{gnuplot@{\em{}gnuplot}!finite-size scaling}
For that purpose you need a file {\tt
  m\_scaling.dat} with three
columns, where the first column contains the system sizes $L$, the second the
values of $p$ and the third contains magnetization $m(p,L)$ for each
data point. First, assume that you know the values for $p_{\rm c},\nu$
and $\beta$. In this case, the actual plot  is done by entering:

\begin{verbatim}
gnuplot> b=0.27
gnuplot> n=1.1
gnuplot> pc=0.222                                    
gnuplot> plot [-1:1] "m_scale.dat" u (($2-pc)*$1**(1/n)):($3*$1**(b/n))
\end{verbatim}
The plot command makes use of the feature that with the u(sing) option
you can transform the data of the input in an arbitrary way. For
each data set, the variables \$1,\$2 and \$3 refer to the first, 
second and third
columns, e.g.\ \$1**(1/n) raises the system size to the power $1/\nu$. 
The resulting plot is shown in Fig.\ \ref{fig:fssGnuplot}. 
Near the transition $p-p_{\rm c}\approx 0$ a good collapse of the data points can be
observed. 

In case you do not know the values of $p_{\rm c},\beta,\nu$ you can start
with some estimated values, perform the plot, resulting probably in a
bad collapse. Then you may alter the parameters iteratively and
watch the resulting changes by plotting again. In this way you can
converge to a set of parameters, where all data points show a
satisfying collapse.

\begin{figure}[ht]
\begin{center}
\scalebox{0.6}{\includegraphics{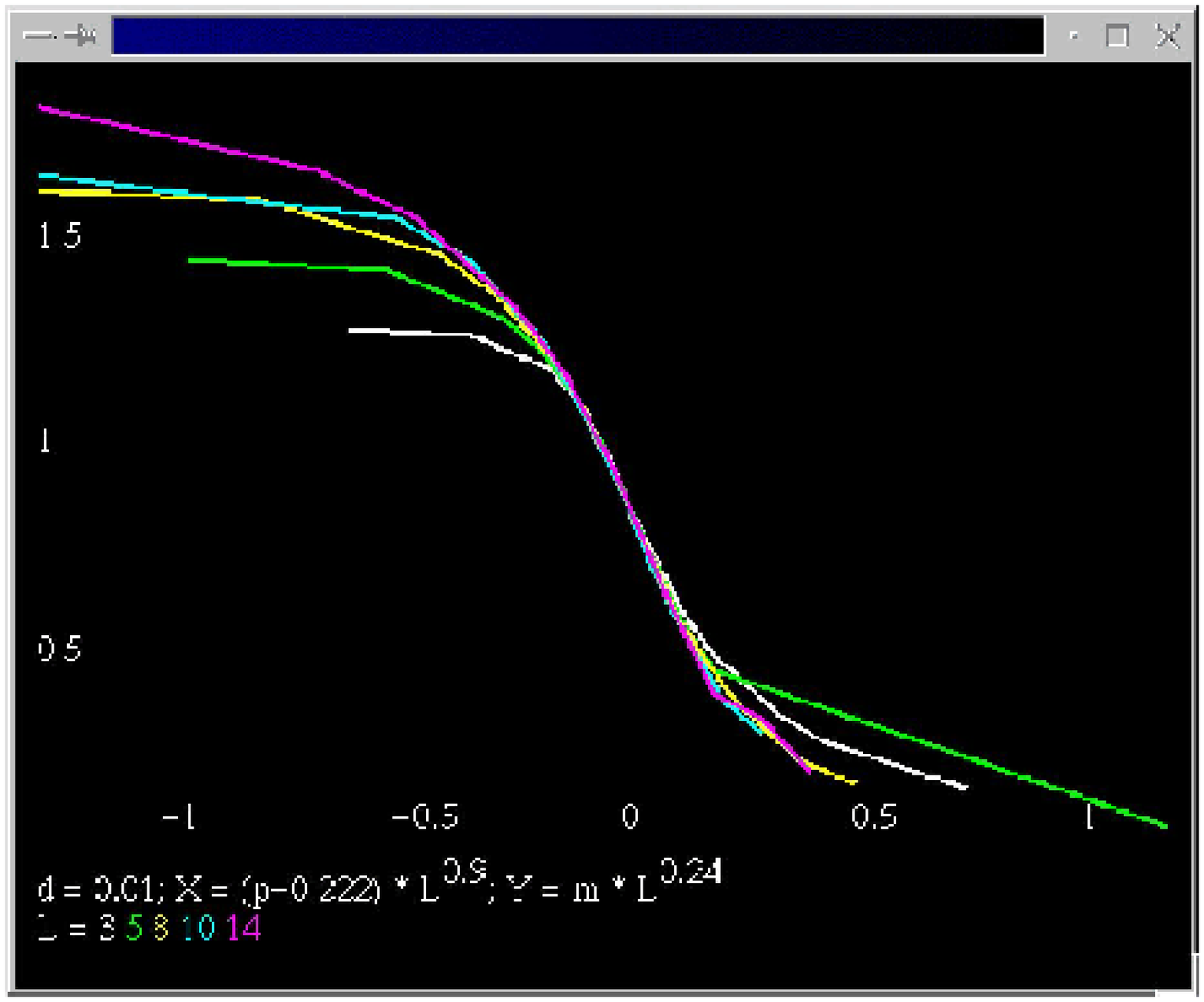}}
\caption{Screen-shot from a window running the {\em fsscale}\/ tool.}
\label{fig:fsscale}
\end{center}
\end{figure}

\index{fsscale@{\em{}fsscale}|(}
The process of determining the finite-size scaling parameters can be
performed more conveniently by using the special purpose program {\em
  fsscale}\/. It can be obtained free of charge from \cite{PRA-fsscale}. 
This tool allows the scaling parameters to be changed 
interactively by pressing buttons on the keyboard, making a
finite-size scaling fit very convenient to perform. Several
different scaling forms are available. To obtain more information,
start the program, with {\tt fsscale -help}. A sample screen-shot is
shown in Fig.\ \ref{fig:fsscale}

Please note that the data have to be presented to {\em fsscale}\/ in a file
containing three columns, where the first column contains the system size,
the second the $x$-value and the third the $y$-value. If you have only
data files with more columns, you can use the standard UNIX tool {\em awk}\/ to
\index{awk} 
project out the relevant columns. For example, assume that your data
file {\tt{}results.dat} has 10 columns, and your are interested in
columns $3,8,$ and 9. Then you have to enter:

\begin{verbatim}
awk '{print $3,$8,$9}' results.dat > projected.dat
\end{verbatim}

You can also use {\em awk}\/ to perform calulations with the values in
the columns, similar to {\em gnuplot}\/, as in

\begin{verbatim}
awk '{print $1+$2, 2.0*$7, $8*$1}' results.dat
\end{verbatim}

\index{data!analysis|)}
\index{finite-size scaling|)}
\index{fsscale@{\em{}fsscale}|)}

\section{Information Retrieval and Publishing}
\label{sec-literature}
In this section some basic information regarding searching for
literature and preparing your own presentations and publications is
given. 

\subsection{Searching for Literature}
\label{sec-litsearch}

Before contributing to the physical community and even publishing
your results, you should be aware of what exists already.  This prevents
you from redoing something which has been done before by someone
else. Furthermore, knowing previous results and many simulation
techniques allows you to conduct your own research projects better.
 Unfortunately, 
much information cannot be found in textbooks. Thus, you must start
to look at the literature. With modern techniques like CD-ROMs and
the Internet this can be achieved very quickly. Within this section, 
it is assumed that you
are familiar with the Internet and are able to use a browser.
In the following list several sources of information are contained.

\begin{itemize}
\item {\bf Your local (university) library}\\ \index{library}
Although the amount of literature is limited from space constraints,
you should always check your local library for suitable textbooks
concerning your area of research. Also many old issues of scientific
journals are yet not available through the Internet, thus you may have to
copy some articles in the library.
\item {\bf Literature databases}\\ \index{literature databases}
In case you want to obtain all articles from a specific author or all
articles on a certain subject, you should consult a literature
database. In physics the {\em INSPEC}\/ \index{INSPEC}
\cite{PRA-inspec} database is the appropriate
source of information. Unfortunately, the access is not free of charge. But
usually your library should allow access to INSPEC, either via CD-ROMS
or via the Internet. If your library/university does not offer an access you
should complain. 

INSPEC frequently surveys almost all scientific journals in the areas of
physics, electronics and computers. For each paper that appears, all
bibliographic information along with the abstract are stored. You
can search the database for example for author names, keywords (in the
abstract or title), publication years or journals. Via INSPEC it is
possible to keep track of recent developments happening in a certain field.

There are many other specialized databases. You should consult the web
page of your library, to find out to which of them you can
access. Modern scientific work is not possible without regularly
checking literature databases.

\item {\bf Preprint server}\\ \index{preprint server}
In the time of the Internet, speed of publication becomes increasingly
important. Meanwhile, many researchers put their publications on the
{\em Los Alamos Preprint server}\/ \cite{PRA-xxx}, 
where they become available world wide 
at most 72 (usually 24) hours after submission. The database is
free of charge
and can be accessed from almost everywhere via a browser. The
preprint database is divided into several sections such as astrophysics
(astro-ph),  condensed matter (cond-mat) or quantum physics
(quant-ph). Similar to a conventional literature database, 
you can search the database, eventually restricted to a
section, for author names, publication years or keywords in the
title/abstract. But furthermore, after you have found an interesting
article, you can download it and print it immediately. File formats
are {\em postscript} \index{postscript@{\em{}postscript}}
and {\em pdf}. \index{pdf@{\em{}pdf}}
The submission can also be in \TeX/\LaTeX\  (see
Sec.\ \ref{sec:publish}).

Please note that there is no editorial processing
 at all, that means you do not have any
guarantee on the quality of a paper. If you like, you can submit a
poem describing the beauty of your garden. Nevertheless, the aim of
the server is to make important scientific results available very
quickly. Thus, before submitting an article, you should be sure that
it is correct and interesting, otherwise you might get a poor reputation.

The preprint server also offers access via email. It is possible to
subscribe to a certain subject. Then every working
day you will receive a list of all new papers which have been 
submitted. This is a very
convenient way of keeping track of recent developments. But be
careful, not everyone submits to the preprint server. Hence, you still
have to read scientific journals regularly.

\item {\bf Scientific journals}\\ \index{scientific journals}
Journals are the most important resources of information in
science. Most of them allow access via the Internet, when your
university or institute has subscribed to them. Some of the most
important physical journals, which are available online, are published
by (in alphabetical order)

\begin{itemize}
\item the American Institute of Physics \cite{PRA-aip}
\item the American Physical Society \cite{PRA-aps}
\item Elsevier Science (Netherlands) \cite{PRA-elsevier}
\item the European Physical Society \cite{PRA-eps}
\item the Institute of Physics (Great Britain) \cite{PRA-iop}
\item Springer Science (Germany) \cite{PRA-springer}
\item Wiley-VCH (USA/Germany) \cite{PRA-wiley}
\item World-Scientific (Singapore) \cite{PRA-world-scientific}
\end{itemize}

\item{\bf Citation databases}\\ \index{citation database}
In every scientific paper some other articles are cited. Sometimes it
is interesting to get the reverse information, i.e.\ 
to obtain all papers which are citing a given article A.
This can be useful, if one wants to learn about the most recent 
developments which are triggered
by article A. In that case you have to access a {\em citation index}\/.
For physics, probably the most important is the 
{\em Science Citation Index}\/ \index{Science Citation Index}
(SCI) \index{SCI} which can be accessed via
the {\em Web of Science}\/ \cite{PRA-web-of-science}. 
You have to ask your system administrator or
your librarian whether and how you can access it from your site.

The {\em American Physical Society} (APS) \index{American Physical Society}
\index{APS} \cite{PRA-aps} also includes
links to citing articles with the online versions of recent papers. If
the citing article is available via the APS as well, you can
immediately access the article from the Internet. This works not only
for citing papers, but also for cited articles.

\item {\bf Phys Net}\\ \index{Phys Net}
If  you want to have access to the web pages of a certain physics
department, you should go via your web browser to the {\em Phys Net}\/
pages \cite{PRA-physnet}. They offer a list of all physics departments in
the world. Additionally, you will find lists of forthcoming 
conferences, job offers
and many other useful links. Also, the home page of
your department probably offers many interesting links to other web pages
related to physics.

\item {\bf Web browsing}\\
Except for the sources mentioned so far, nowadays much information
is available on net. Many researchers present their work, their
results and their publications on their home pages. Quite often
talks or computer codes can also be downloaded.

In case you cannot find a specific page through the {\em Phys Net}\/ (see
above), or you are interested in obtaining all web pages concerning a
specific subject, you should ask a {\em search
  engine}\/. \index{search!engine} There are some
very popular all purpose engines like {\em Yahoo}\/ \index{Yahoo}
\cite{PRA-yahoo} or {\em Alta Vista}\/ \index{Alta Vista}
\cite{PRA-alta-vista}. A very convenient way to start a query on several
search engines in parallel is a {\em meta search engine}\/, 
e.g.\ {\em Metacrawler}\/ \cite{PRA-metacrawler}. 
To find out more, please contact a search engine.
\end{itemize}

\subsection{Preparing Publications} 
\label{sec:publish}

In this section tools for two types of presenting your results are
covered: via an article/report or in a talk. For writing papers, it is
recommended that you use {\em \TeX/\LaTeX}\/. 
Data plots can be produced using the programs
explained in the last section. For drawing figures and making
transparencies, the program {\em xfig}\/ offers a large
functionality. To create three-dimensional perspective images, the
program {\em Povray}\/ can be used. \LaTeX, {\em xfig}\/ and {\em
  Povray}\/ are 
introduced in this section.

\index{TeX@\TeX} \index{LateX@\LaTeX|(}
First, \TeX/\LaTeX\  is explained. It is a typesetting system rather than
a word processor. The basic program is \TeX, \LaTeX\  is an extension to
facilitate the application. In the area of theoretical computer science,
the combination of \TeX\  and \LaTeX\  is a
widespread standard. When submitting an article electronically to a
scientific journal usually \LaTeX\  has to be used.
Unlike the conventional office packages, with \LaTeX\  you do not see
the text in the form it will be printed, i.e.\ \LaTeX\  is not a WYSIWYG
(``What you see is what you get'') program. The text is entered in a
conventional text editor (like {\em Emacs}\/) and all formatting is done via
special commands. An introduction to the \LaTeX\  language can be found
e.g.\ in Refs.\ \cite{PRA-lamport1994,PRA-tug}. 
Although you have to learn several
commands, the use of \LaTeX\  has several advantages:
\begin{itemize}
\item The quality of the typesetting is excellent. It is much better than
  self-made formats. You do not have to care about the layout. But
  still, you are free to change everything according to your requirements.
\item Large projects do not give rise to any problems, in contrast to
  many commercial office programs. When treating a \LaTeX\  text, your 
computer will
  never complain when your text is more than 300 pages or contains many
  huge post-script figures.
\item Type setting of formulae is very convenient and fast. You do not have to
  care about sizes of indices of indices etc. Furthermore, in case you
  want for example to replace all $\alpha$ in your formulae with
  $\beta$, this can be done with a conventional replace, by replacing
  all {\verb! \alpha!} strings by a {\verb! \beta!} strings. For the case
  of an office system, please do not ask how to do this conveniently.
\item There are many additional packages for enhanced styles such as
  letters, transparencies or books. The {\em bibtex}\/
  package is very convenient, which allows a nice literature database
  to be build  up.
\item Since you can use a conventional editor, the writing process is
  very fast. You do not have to wait for a huge packet to come up.
\item On the other hand, if you still prefer a WYSIWYG (``what you see
  is what you get'') system, there
  is a program called {\em lyx}\/ \cite{PRA-lyx} which operates like a
  conventional word processor but creates \LaTeX\  files as
  output. Nevertheless, once you get used to \LaTeX, you will never
  want to loose it. 
\end{itemize}

Please note that this text was written entirely with \LaTeX.
Since \LaTeX\  is a type setting language,  you have to compile your text
to create the actual output. Now, an example is given of what a \LaTeX\  text
looks like and how it can be compiled. This example will just give you
an impression of how the system operates. 
For a complete reference, please consult the literature mentioned above.
 
The following file {\tt example.tex} 
produces a text with different fonts and a formula:
\begin{verbatim}
\documentclass[12pt]{article}
\begin{document}
This is just a small sample text. You can write some words {\em
emphasized}\/, or in {\bf bold face}. Also different {\small sizes} 
are possible. 

An empty line generates a new paragraph. \LaTeX\  is very convenient 
for writing formulae, e.g.
\begin{equation}
M_i(t) = \frac{1}{L^3} \int_V x_i \rho(\vec{x},t) d^3\vec{x}
\end{equation}
\end{document}
\end{verbatim}

The first line introduces the type of the text ({\tt article}, which is
the standard) and the font size. You should note that all tex commands
begin with a backslash ($\backslash$), in case you want to write a backslash in
your text, you have to enter {\verb! $\backslash$!}. 
The actual text is written between
the lines starting with {\tt $\backslash$begin\{document\}} and ending with {\tt
  $\backslash$end\{document\}}. You can observe some commands such as 
{\tt $\backslash$em},
{\tt $\backslash$bf} or {\tt $\backslash$small}. 
The {\tt \{ \} } braces are used to mark
blocks of text. Mathematical formulae can be written
e.g.\ with {\tt $\backslash${}begin\{equation\}} and 
{\tt $\backslash${}end\{equation\}}. For
the mathematical mode a huge number of commands exists. Here only
examples for Greek letters ({\tt $\backslash${}alpha}), 
subscripts ({\tt x\_i}),
fractions ({\tt $\backslash${}frac}), integrals ({\tt
  $\backslash${}int}) and vectors ({\tt$\backslash$vec}) are given.

The text can be compiled by entering {\tt latex example.tex}. This is
the command for UNIX, but \LaTeX\  exists for all operating systems. Please
consult the documentation of your local installation.

The output of the compiling process is the file {\tt example.dvi}, where
``dvi'' means ``device independent''. The {\tt .dvi} file can be
inspected on screen by a viewer via entering {\tt xdvi example.dvi} or
converted into a postscript \index{postscript@{\em{}postscript}}
file via typing 
{\tt dvips -o example.ps example.dvi}
and then transferred to a printer. On many systems it can be printed
directly as well. The result will look like this:
\begin{quote}
This is just a small sample text. You can write some words {\em
  emphasized}\/, or in {\bf bold face}. Also different {\small sizes} are
possible. 

An empty line generates a new paragraph. \LaTeX\  is very convenient for
writing formulae, e.g.
\begin{equation}
M_i(t) = \frac{1}{L^3} \int_V x_i \rho(\vec{x},t) d^3\vec{x}
\end{equation}
\end{quote}

This example should be sufficient to give you an impression of what the
philosophy of \LaTeX\  is. Comprehensive instructions are beyond the
scope of this section, please consult the literature
\cite{PRA-lamport1994,PRA-tug}.

\begin{figure}[h!]
\begin{center}
\scalebox{0.6}{\includegraphics{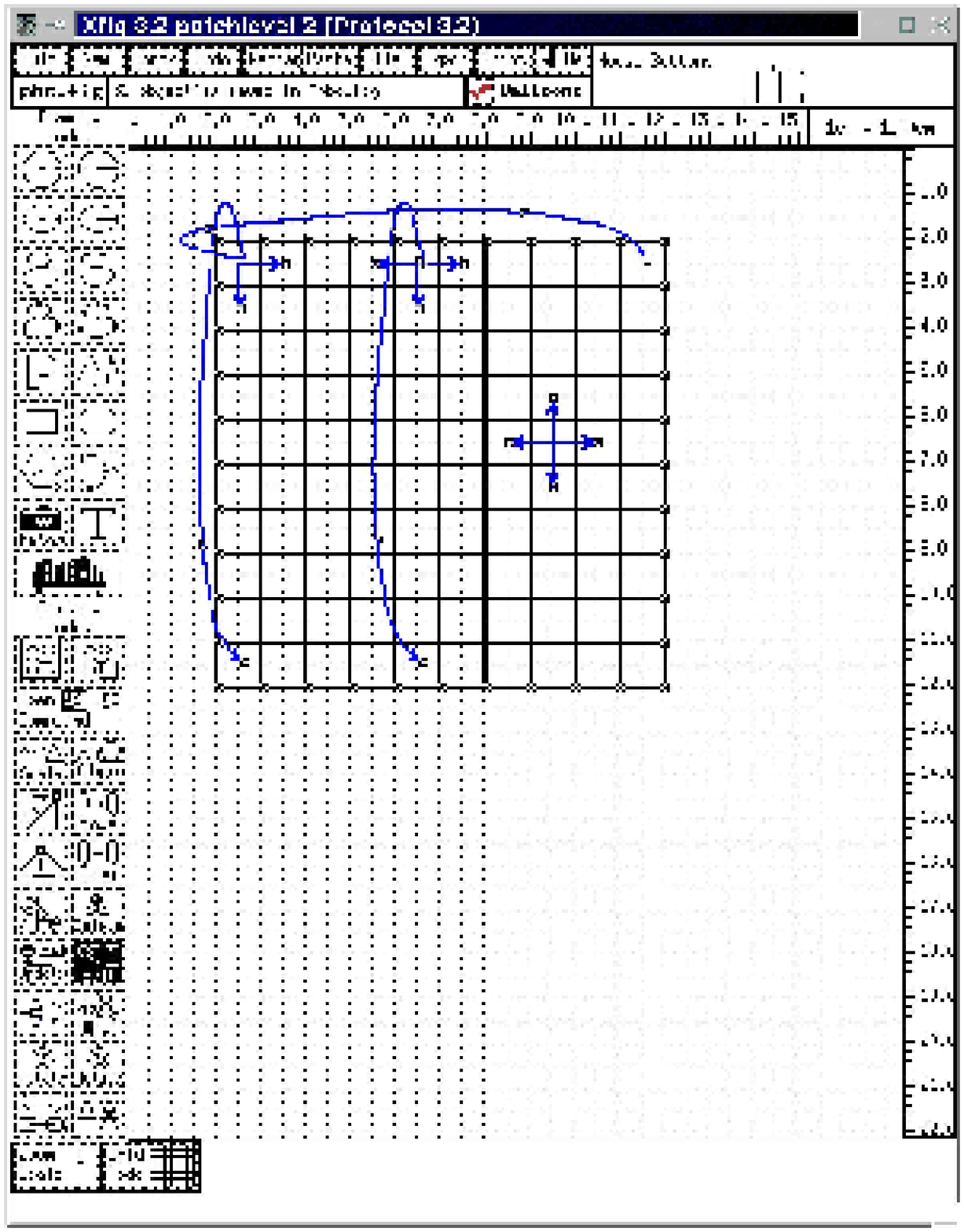}}
\caption{A sample screen-shot showing the {\em xfig}\/ program.}
\label{fig:xfig:sample}
\end{center}
\end{figure}
\index{LateX@\LaTeX|)}

Under UNIX/Linux, the spell checker {\em{}ispell}\/ \index{ispell}
is available. It
allows a simple spell check to be performed. The tool is built on a dictionary,
i.e.\ a huge list of
known words. The program scans any given text, also a special \LaTeX\ 
mode is available. Every time a  word occurs, which is not contained in the
list, {\em ispell}\/ stops. Should similar words exist in the
list, they are suggested. Now the user has to decide whether the word
should be replaced, changed, accepted or even added to the
dictionary. The whole text is treated in this way. Please note that
many mistakes cannot be found in this way, especially when the misspelled
word is equal to another word in the dictionary. However, at least {\em
  ispell}\/ finds many spelling mistakes quickly and conveniently, so
you should use the tool.

\index{xfig@{\em{}xfig}|(}
Most scientific texts do not only contain text, formulae and curves,
but also schematic figures showing the models, algorithms or devices
covered in the publication. A very convenient but also simple tool to
create such figures is {\em xfig}\/. It is a window based
vector-oriented drawing program. Among its features are the creation
of simple objects like lines, arrows,  polylines, splines, arcs 
as well as rectangles, 
circles and other closed, possibly filled, areas. Furthermore you can
create text or include arbitrary (eps, jpg) pictures files. 
You may place the objects on different layers which allows 
complex sceneries to be created. 
Different simple objects can be combined into more
complex objects. For editing you can move, copy, delete, rotate or
scale objects. To give you an impression what {\em xfig}\/ looks like, in
Fig.\ \ref{fig:xfig:sample} a screen-shot is shown, displaying {\em xfig}\/
with the picture that is shown in Fig.\ \ref{fig:pbc}. Again, for
further help, please consult the online help function or the {\em man}\/ pages.

The figures can be saved in the internal fig format, and exported in
several file formats such as (encapsulated) {\em postscript}\/, 
\index{postscript@{\em{}postscript}}
\LaTeX, {\em Jpeg}\/, \index{jpeg@{\em{}Jpeg}}
{\em Tiff}\/ \index{tiff@{\em{}Tiff}} or bitmap.\index{bitmap}
 The {\em xfig}\/ program can be called in a way that it
produces just an output file with a given fig input file. This is very
convenient when you have larger projects where some small picture objects
are contained in other pictures and you want to change the appearance
of the small objects in all other files. With the help of the {\em
  make}\/ program pretty large projects can be realized.

Also, {\em xfig}\/ is very convenient when creating
 transparencies for talks, which
is the standard method of presenting results in physics. With different
colors, text sizes and all the objects mentioned before, 
very clear transparencies can be created
quickly. The possibility of including
picture files, like {\em postscript}\/ 
\index{postscript@{\em{}postscript}}
files which were created by a data plotting
program such as {\em xmgr}\/, is very helpful. 
In the beginning it may seem that more effort is necessary
than when creating the transparencies by hand. However, once you have a
solid base of transparencies you can reuse many parts and preparing a
talk may become a question of minutes. In particular, when your
handwriting looks awful, the audience will be much obliged for
transparencies prepared with {\em xfig}\/.

Last but not least, please note that {\em xfig}\/ is vector oriented,
but not pixel oriented. Therefore, you cannot treat pictures like jpg files
(e.g.\ photos) and
apply operations like smoothing, sharpening or filtering. For these
purposes the package {\em gimp}\/ is suitable. It is freely available
again from GNU \cite{PRA-gnu}.

It is also possible to draw three-dimensional figures with {\em xfig}\/, but
there is no special support for it. This means, {\em xfig}\/ has only a
two-dimensional coordinate system. \index{xfig@{\em{}xfig}|)}
A very convenient and powerful 
tool for making
three-dimensional figures is {\em Povray}\/ \index{Povray@{\em{}Povray}|(}
(Persistence Of Vision
RAYtraycer).  Here, again, only a short
 example is given, for a detailed documentation please
refer to the home page \cite{PRA-povray}, where the program can be
downloaded for many operating systems free of charge.

{\em Povray}\/ is, as can be realized from its name, a {\em
  raytracer}\/. \index{raytracer} This
means you present a scene consisting of several objects to the program.
These objects have characteristics like color, reflectivity or
transparency. Furthermore the position of one or several light sources
and a virtual camera have to be defined. The output of a raytracer is a
photo-realistic 
picture of the scene, seen through the camera. The name ``raytracer''
originates from the fact that the program creates a picture by
starting several rays of light
at the light sources and traces their way through the scene, where they
may be absorbed, reflected or refracted, until
they hit the camera, disappear into infinity or become too weak. Hence,
the creation of a picture may take a while, depending on the
complexity of the scene.

A scene is described in a human readable file, it can be entered with
any text editor. But for more complex scenes, special editors exist,
which allow a scene  to be created interactively. Also several tools for
making animations are available on the Internet. Here, a simple
example is given. The scene consists of three spheres connected by two
cylinders, forming a molecule. Furthermore, a light source, a camera,
an infinite plane and the background color are defined. Please note
that a sphere is defined by its center and a radius and a cylinder by
two end points and a radius. Additionally, for all objects color
information has to be included, the center sphere is slightly
transparent.  The scene description file {\tt
  test1.pov} reads as follows:
\hfill
\begin{verbatim}
#include "colors.inc"  

background { color White }

sphere {  <10, 2, 0>, 2
    pigment { Blue } }

cylinder { <10, 2, 0>,  <0, 2, 10>, 0.7         
    pigment { color Red } } 

sphere {  <0, 2, 10>, 4
    pigment { Green transmit 0.4} }

cylinder { <0, 2, 10>,  <-10, 2, 0>, 0.7         
    pigment { Red } } 

sphere {  <-10, 2, 0>, 2
    pigment { Blue } }

plane { <0, 1, 0>, -5
    pigment { checker color White, color Black}}

light_source { <10, 30, -3> color White} 

camera {location <0, 8, -20>
    look_at  <0, 2,  10>
    aperture 0.4}
\end{verbatim}

\begin{figure}[ht]
\begin{center}
\scalebox{0.6}{\includegraphics{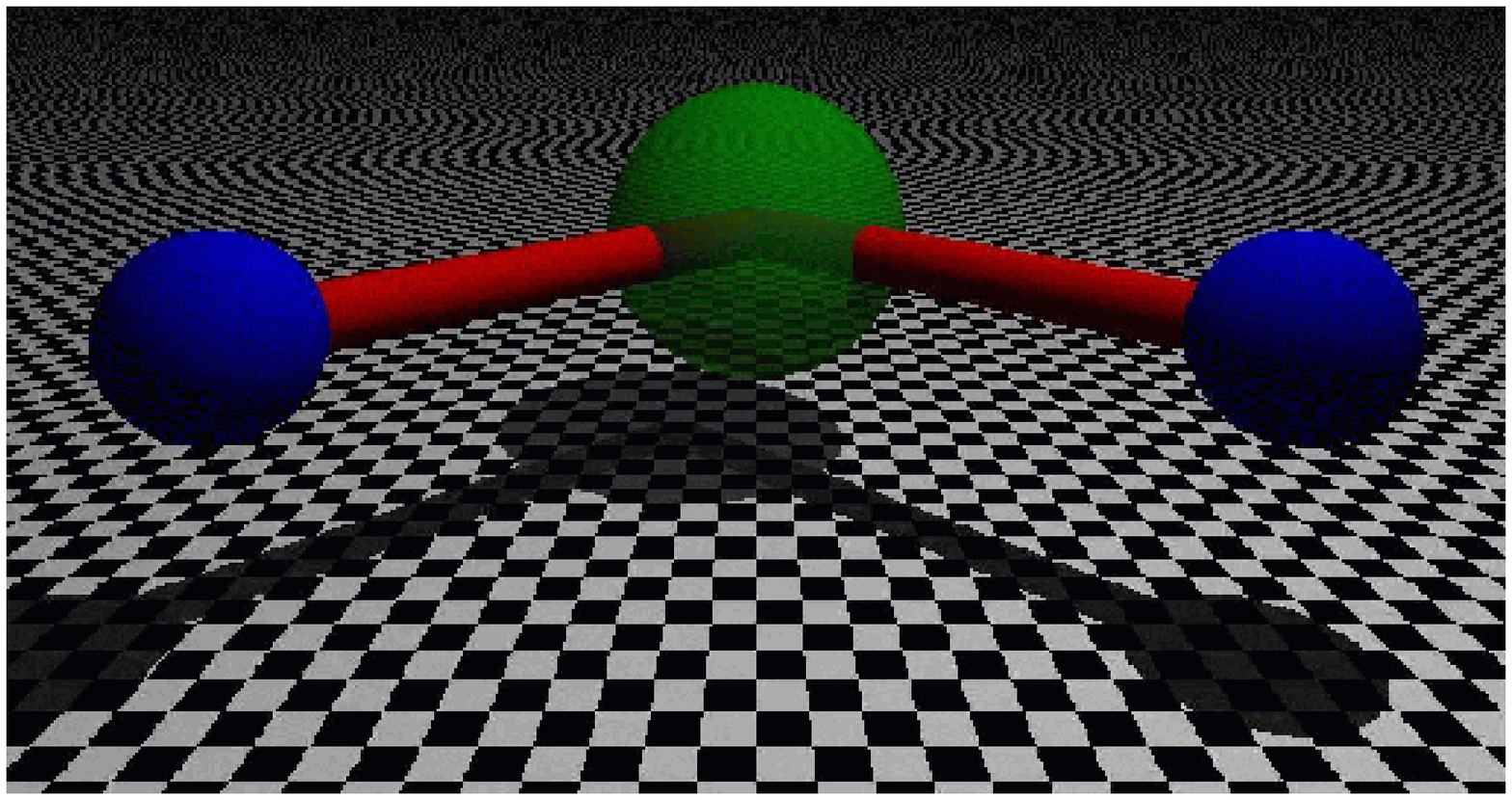}}
\caption{A sample scene created with {\em Povray}\/.}
\label{fig:pov:sample}
\end{center}
\end{figure}

The creation of the picture is started by calling (here on a Linux
system via command line) {\tt x-povray +I test1.pov}. The resulting
picture is shown in Fig.\ \ref{fig:pov:sample}, please note the
shadows on the plane.

{\em Povray}\/ is really powerful. You can create almost arbitrarily shaped
objects, combine them into complex objects and impose many
transformations. Also special effects like blurring or fog are
available. All features of {\em Povray}\/ are described in a 400 page manual.
The use of {\em Povray}\/ is widespread in the artists community. For
scientists it is very convenient as well, because you can easily convert e.g.
configuration files of molecules or three-dimensional domains of
magnetic systems into nice
looking perspective pictures. 
This can be accomplished by writing a small program which
reads e.g your configuration file containing a list of positions of
atoms and a
list of links, and puts for every atom a sphere and for every link a
cylinder into a {\em Povray}\/ scene file. Finally the program must add suitable
chosen light sources and a camera. Then, a three-dimensional
pictures is created by calling {\em Povray}\/.
\index{Povray@{\em{}Povray}|)}
 
The tools described in this section, should allow all technical
problems occurring in the process of preparing a publication (a
``paper'') to be solved. Once you
have prepared it, you should give it  to at least one other person, who
should read it carefully. Probably he/she will find some errors or
indicate passages which might be difficult to understand or
that are misleading. You should always take such comments very seriously,
because the average reader knows much less about your problem than you do.

After all necessary changes have been performed, and you and other
readers are satisfied with the publication, you can submit it to
a scientific journal. You should choose a journal which suits your
paper. Where to submit, you should discuss with experienced
researchers. It is not possible to give general advice on this issue. 
Nevertheless, technically the submission can be performed nowadays
almost everywhere electronically. For a list of publishers of some important
journals in physics, please see Sec.\ \ref{sec-litsearch}. Submitting
 one paper to several journals in parallel is not
allowed. However, you should
also consider  submitting  to the preprint server
\cite{PRA-inspec} as well to make your results quickly available to the physics
community. 

Nevertheless, although this text provides many useful hints
concerning performing computer simulations,
the main part of the work is still having good
ideas and carefully conducting the actual research projects.  

%\begin{thebibliography}{99}

\end{document}